\documentclass[preprintnumbers,superscriptaddress,nofootinbib,aps,prd,floatfix,twocolumn]{revtex4-1}

\pdfoutput=1
\usepackage{graphicx}
\usepackage{latexsym}
\usepackage{amsfonts}
\usepackage{amssymb}
\usepackage{amsmath}
\usepackage{slashed}
\usepackage{feynmp}
\usepackage{hyperref}
\usepackage{url}
\usepackage{color}
\usepackage{cancel}
\usepackage{multirow,array}

\usepackage[caption=false,margin=2ex]{subfig}
\usepackage{ragged2e}
\DeclareCaptionJustification{justified}{\justifying}

\begin{document}
\date{\today}

\preprint{
\begin{minipage}{5cm}
\small
\flushright
UCI-HEP-TR-2022-19
\end{minipage}}

\title{Directional Neutrino Searches for Galactic Center Dark Matter at Large Underground LArTPCs}

\author{Matthew R.~Buckley}
\affiliation{Department of Physics and Astronomy, Rutgers University, Piscataway, NJ 08854, USA}

\author{Andrew Mastbaum}
\affiliation{Department of Physics and Astronomy, Rutgers University, Piscataway, NJ 08854, USA}

\author{Gopolang Mohlabeng}
\affiliation{Physics Department, Brookhaven National Laboratory, Upton, New York 11973, USA}
\affiliation{The McDonald Institute and Department of Physics, Queen's University, Kingston, Ontario K7L 3N6, Canada}
\affiliation{Perimeter Institute for Theoretical Physics, Waterloo, Ontario N2L 2Y5, Canada}
\affiliation{Department of Physics and Astronomy, University of California, Irvine, Irvine, CA 92697, USA}

\begin{abstract}
We investigate the sensitivity of a large, underground LArTPC-based neutrino detector to dark matter in the Galactic Center annihilating into neutrinos.
Such a detector could have the ability to resolve the direction of the electron in a neutrino scattering
event, and thus to infer information about the source direction for individual neutrino events.
We consider the improvements on the expected experimental sensitivity that this directional
information would provide. Even without directional information, we find a DUNE-like LArTPC detector
is capable of setting limits on dark matter annihilation to neutrinos for dark matter masses above
30~MeV that are competitive with or exceed current experimental reach. 
While currently-demonstrated angular resolution for low-energy electrons is insufficient to allow any significant
increase in sensitivity, these techniques could benefit from improvements to algorithms and the
additional spatial information provided by novel 3D charge imaging approaches. We consider the impact of
such enhancements to the resolution for electron directionality, and find that where
electron-scattering events  can be distinguished from charged-current neutrino interactions,
limits on dark matter annihilation in the mass range where solar neutrino backgrounds 
dominate ($\lesssim 15$~MeV) can be significantly improved using directional information, and would be competitive with existing limits using $40$~kton$\times$year of exposure.
\end{abstract}

\maketitle 

\section{Introduction}
The existence of dark matter is well-attested through astrophysical and cosmological probes \cite{Bertone:2016nfn,ParticleDataGroup:2016lqr}, but its particle nature remains completely unknown. This lack of knowledge requires a multi-pronged experimental approach to cover as much of the theory space as possible. A critical segment of this experimental program is indirect detection of dark matter through its annihilation or decay into Standard Model particles in the Universe today which are then seen by Earth- and space-based observatories. Gamma-ray, X-ray, and radio telescopes can constrain dark matter annihilation or decay either directly into photons or into states carrying electric or color charges that generate photons through either decay or synchrotron radiation. Ref.~\cite{Leane:2020liq} provides a recent comprehensive review of indirect detection constraints.

Most difficult to constrain are dark matter matter couplings to neutrinos \cite{Beacom:2006tt,Yuksel:2007ac}, which could perhaps be the main channel through which dark matter interacts with the visible particles \cite{Olivares-DelCampo:2017feq,Blennow:2019fhy}. 
A direct coupling between neutrinos and dark matter may generate the small neutrino masses \cite{Boehm:2006mi,Falkowski:2009yz,Cosme:2020mck, Davoudiasl:2018hjw}, and provide a mechanism to obtain the correct dark matter relic abundance in the early Universe \cite{Berlin:2018ztp,Du:2020avz, Chao:2020bti}. A dark matter-neutrino interaction could also affect cosmology by leaving an imprint on the cosmic microwave background (CMB) \cite{Mangano:2006mp, Serra:2009uu} and structure formation \cite{Wilkinson:2014ksa,Bertoni:2014mva,DiValentino:2017oaw}. Finally, at late times, high energy extra-galactic neutrinos scattering off dark matter halos could lead to an attenuation of the neutrino flux on Earth \cite{Reynoso:2016hjr,Arguelles:2017atb,Choi:2019ixb}.

Dark matter annihilation to charged or strongly coupled unstable Standard Model particles (e.g., 2$^{\rm nd}$ or $3^{\rm rd}$ generation quarks, or $W/Z$ bosons) can generate neutrinos in the subsequent cascade decays. Alternatively, dark matter may annihilate directly into a monoenergetic neutrino pair (a purely invisible channel) with energy equal to the dark matter mass, i.e.,  
$E_{\nu} = m_{\chi}$. In this paper, we will consider this latter possibility.

Due to their elusive nature and non-trivial backgrounds, indirect detection of
neutrino final states faces significant barriers, especially in the low-mass
regime ($m_\chi = E_\nu \sim {\cal O}(10)~{\rm MeV}$) where the neutrino interaction cross section is small and the solar
neutrino background is large. The importance of indirect searches for dark matter annihilations into neutrinos has long been
recognized \cite{Palomares-Ruiz:2007trf,Klop:2018ltd,McKeen:2018pbb,Arguelles:2019ouk}, and existing limits in this energy regime have been established using data from
Borexino \cite{2011PhLB..696..191B}, KamLAND \cite{KamLAND:2011bnd,KamLAND:2021gvi}, and Super-Kamiokande 
\cite{Super-Kamiokande:2015qek}. IceCube provides complementary constraints at higher masses and neutrino energies \cite{IceCube:2015mgt,IceCube:2016umi}. We refer to Ref.~\cite{Arguelles:2019ouk} for a phenomenological reanalysis
of the neutrino data in terms of dark matter annihilation, which also provides an up-to-date review of existing constraints.

The dark matter annihilation rate is proportional to the density squared along the line-of-sight. As a result, the strongest astrophysical source of a neutrino indirect signal would appear to originate from the highest concentration of dark matter near the Earth: the Galactic Center. This means there is significant directionality in the neutrino signal, which can be a powerful tool in disentangling signal and background.

For dark matter producing neutrinos in the $5-100$~MeV energy range, large-scale
Liquid Argon Time Projection Chambers (LArTPCs) present an exciting opportunity for improved constraints \cite{Rott:2016mzs,Klop:2018ltd,Rott:2019stu,Asai:2020qlp,Kelly:2019wow,DUNE:2021gbm}.
This technology offers excellent prospects for
large exposure and detailed imaging of neutrino interactions. In particular,
potential for direction reconstruction of low-energy neutrino scatters, and
the possibility to discriminate neutrino-nucleus and neutrino-electron
interactions by tagging correlated final state activity makes this an
attractive approach for astrophysical neutrino sources \cite{Moller:2018kpn,Castiglioni:2020tsu,Q-Pix:2022zjm}. As a benchmark to evaluate the performance of future
large, underground LArTPCs, in this work we consider the indirect detection reach of a detector similar to the DUNE far detector \cite{DUNE:2020lwj, DUNE:2020ypp}. 
We are in particular interested in whether the ability of such a LArTPC to afford some information 
about the directionality of the incoming neutrino can be used to increase the
sensitivity of indirect detection searches by reducing backgrounds.

We evaluate the potential sensitivity of this benchmark experiment
to dark matter annihilation into neutrinos, both with and without directional
capabilities. Even without directional information, we show that competitive experimental limits can be achieved with 40~kton$\times$year of exposure for dark
matter in the mass range $5-50$~MeV. The lower end of this range is set by an assumed energy threshold, in consideration of triggering and low-energy radiological backgrounds. Above $\sim 30$~MeV, the projected sensitivity would out-perform all existing limits.

Below $\sim 20$~MeV where solar neutrino backgrounds dominate, directional information can significantly improve sensitivity.
This improvement can be enhanced (up to an order of magnitude improvement) if efficient event-by-event discrimination between neutrino-nucleus and neutrino-electron scattering can be achieved. If event discrimination and directional capabilities are demonstrated in a LArTPC, such an experiment could provide strong constraints on dark matter annihilation into neutrinos
with masses below $20$~MeV, in addition to the record-setting limits at higher masses.

We describe the experimental concept in Section~\ref{sec:expt}, including our
assumptions for the resolution, directional capabilities, and
ability to distinguish scattering events by their final states.
Section~\ref{sec:signals} outlines the expected signal from dark matter
annihilation in the Galactic Center, convolved with the assumed detector response. In Section~\ref{sec:backgrounds}, we describe the main
backgrounds, which can be divided into two main categories: isotropic
backgrounds and Solar backgrounds. Our statistical approach resulting in
model-independent limits is described in Section~\ref{sec:limits}, which we
cast as limits on a gauged $L_\mu-L_\tau$ model in Section~\ref{sec:interpretations}. 

\section{Experiment Description \label{sec:expt}}

As a model for a large, deep underground LArTPC-based neutrino experiment,
we consider the parameters and expected performance of
the DUNE Far Detector \cite{DUNE:2020lwj, DUNE:2020ypp} as a concrete benchmark.
Our assumed detector is thus a Liquid Argon Time Projection Chamber (LArTPC)
with a 13.7~kilotonne active liquid argon volume. In such a detector, final state
charged particles are detected through their ionization of the bulk argon: a
uniform electric field drifts these ionization electrons to an anode plane where
signals are detected and digitized. In a traditional wire-readout LArTPC,
charge sensing is performed using an array of parallel planes of wires, typically three
planes with spacing between planes and between wires approximately 3--5~mm,
yielding three projections that are combined with timing to form a 3D image of
the deposited charge.
Planned next-generation LArTPC detectors such as the DUNE Liquid Argon Near
Detector (ND-LAr) \cite{DUNE:2021tad} will instead employ novel pixel-based readout systems
\cite{Asaadi:2019kof, Dwyer:2018phu} to instrument the
anode plane with charge-sensitive 2D ``pixel'' pads, providing 3D information without
requiring inter-plane matching. This approach leads to reduced ambiguities and lower noise, yielding improved spatial resolution for low-energy signals and a more uniform response across track angles relative to the readout plane. With significant benefits to direction reconstruction for MeV-scale neutrino signals, this is a promising technology for future large-scale, deep-underground LArTPCs.

The sensitivities we estimate are based on
a nominal year-long exposure for a four-module detector, noting that
this exposure may be obtained with longer running in a subset of modules, and extended with a longer run time.
In order to achieve 40~kton$\times$year exposure in roughly this calendar time, it is
assumed that events of interest will be recorded continuously, not only when an accelerator
neutrino beam is firing. This may be achieved in a conventional wire-based readout
LArTPC using a compressed, zero-suppressed data stream as has been developed for
supernova burst sensitivity \cite{MicroBooNE:2020mqg}, or using a pixel-based readout scheme wherein
self-triggering 2D pixel channels naturally yield continuous readout with low
data volume \cite{Dwyer:2018phu, Nygren:2018rbl}.

The signals of interest in the LArTPC are final state charged particles produced in
the interactions of neutrinos with argon nuclei and electrons. Of particular interest
are $\nu_e+{^{40}\mathrm{Ar}}$ charged-current (CC) scattering and $\nu+e^-$ elastic scattering (ES),
both of which produce an energetic final state electron. In this work, we are focused on
neutrinos with energies below 50~MeV. In this regime, particularly below $E_\nu\lesssim30$~MeV,
the $\nu_e$ charged-current interactions are dominated by transitions to low-lying nuclear excited states,
with a subsequent de-excitation producing a cascade of $\gamma$-rays, which go on to
deposit energy nearby via electron Compton scattering. Elastic scattering (ES) --- which proceeds for
all flavors via neutral current channels, but with an enhanced cross section for $\nu_e$ due
to an additional charged-current ES contribution --- yields an energetic final state electron
with no excited nucleus, and thus no associated $\gamma$ cascade. This provides a means for
discriminating between CC and ES events, which has been explored in the context
of supernova neutrino detection in Ref.~\cite{Castiglioni:2020tsu}.
We will apply this ability to distinguish between the ES and CC events later in our analysis.

\subsection{Directional Reconstruction}
In this work, we are interested in the sensitivity to the direction of flight of a neutrino which produces a scattered electron in the detector. This information can only be imperfectly reconstructed as the direction of the scattered electron is not fully aligned with the unseen neutrino's path. This incomplete correlation means that, even if the experiment could reconstruct the electron direction with  perfect angular resolution, the neutrino direction would only be imperfectly determined. 

\begin{figure}[htp]
\includegraphics[width=0.9\columnwidth]{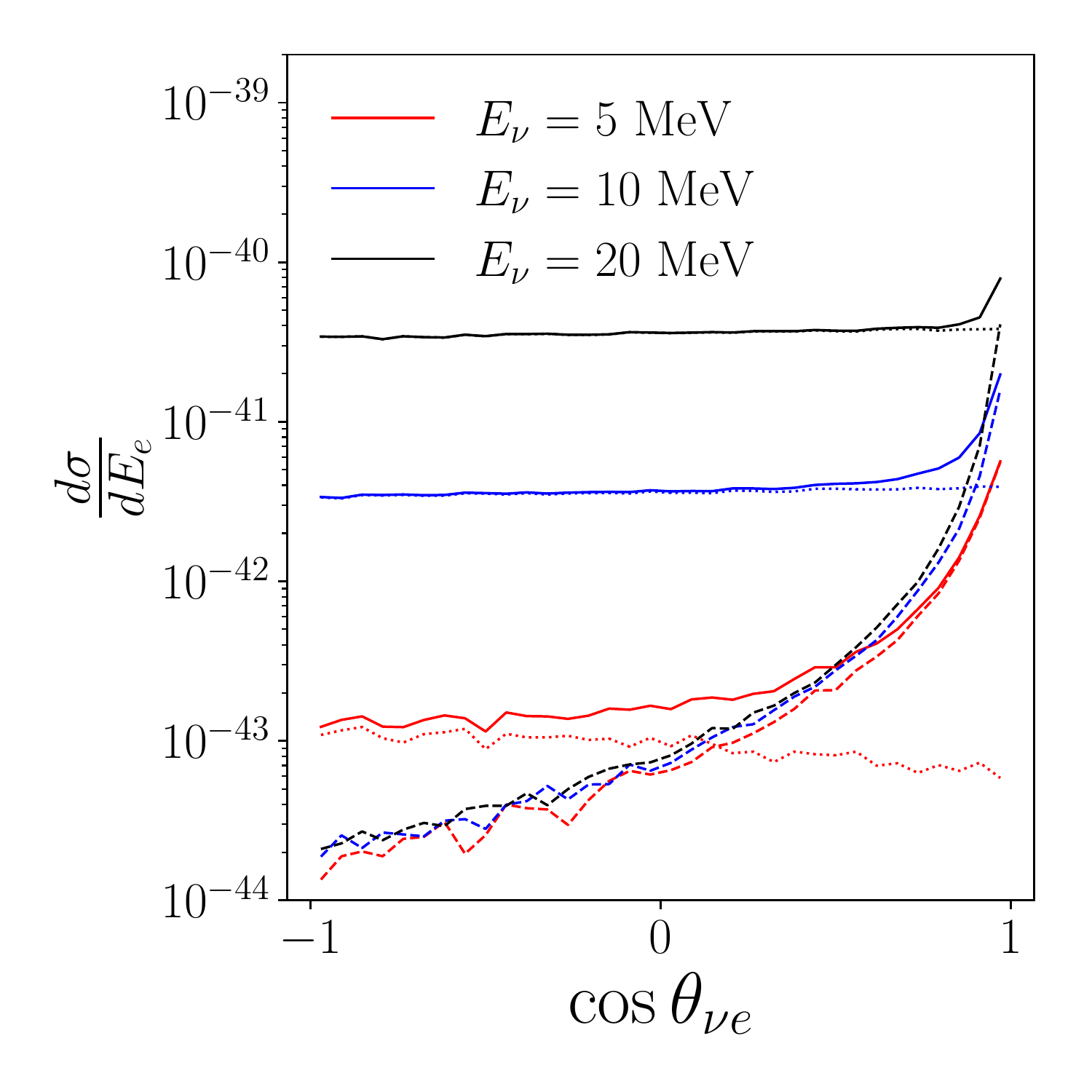}\\
\includegraphics[width=0.9\columnwidth]{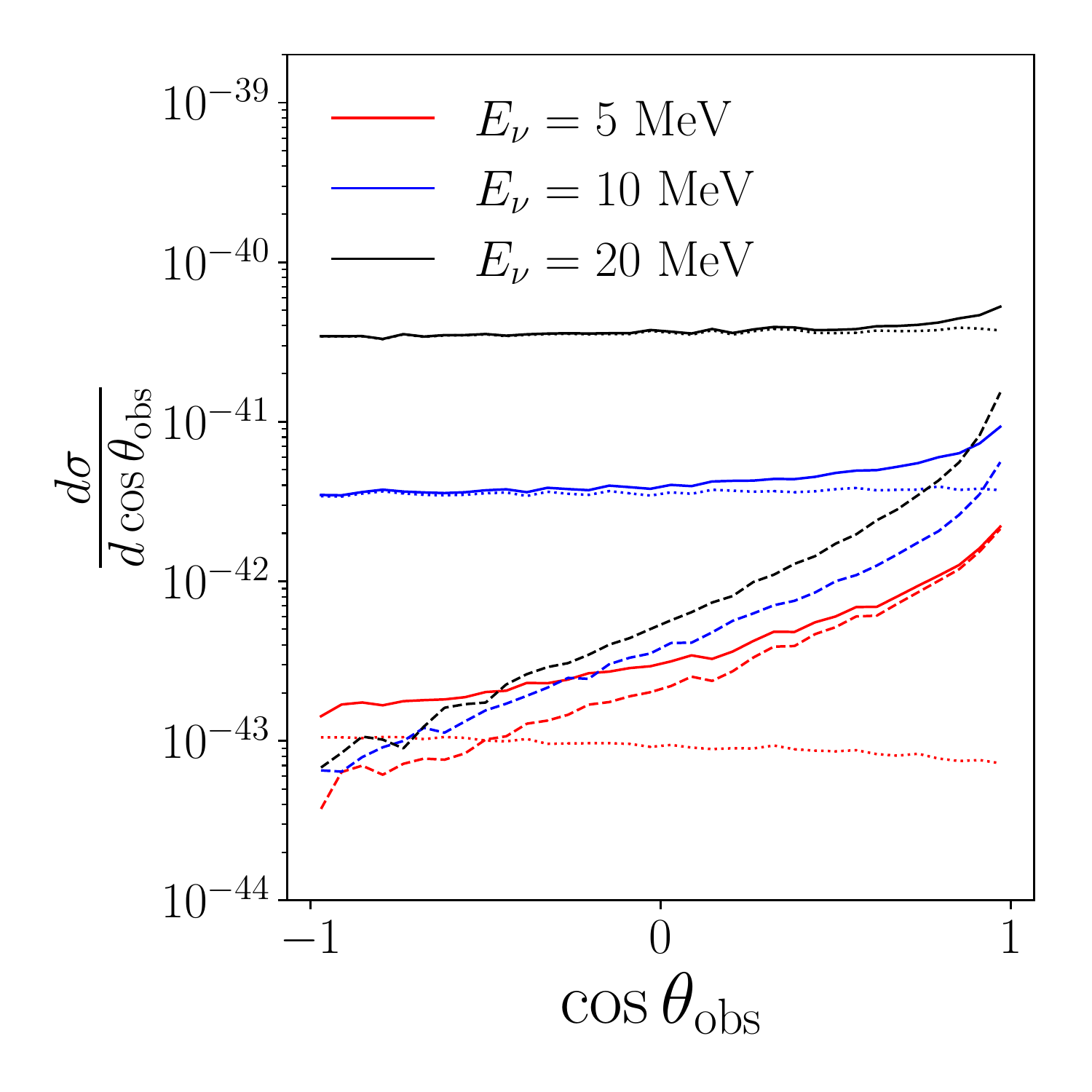} 
\caption{{\it Top:} Differential cross section for neutrino scattering per argon atom via electron-scattering (ES, dashed) \cite{Giunti:2007ry}, charged-current (CC, dotted) \cite{marleyCPC,marleyPRC,MARLEYv1.2.0}, and total (solid) as a function of the recoil angle $\theta_{\nu e}$ of the electron relative to the initial neutrino direction, for 5~MeV (red), 10~MeV (blue), and 20~MeV (black) neutrino energies. {\it Bottom:} Differential cross section for observed direction of the scattered electron assuming a detector angular resolution of $30^\circ$, with labeling as in left panel. Distributions were calculated using Monte Carlo simulations from MARLEY.
\label{fig:dsigmadtheta}}
\end{figure}
We consider the cases of electrons
produced either in $\nu_e-{^{40}\mathrm{Ar}}$
charged-current (CC) interactions
or in neutrino-electron elastic scattering (ES). In the latter case, the
direction of the final state electron is strongly correlated with the
incoming neutrino. In the first panel of Figure~\ref{fig:dsigmadtheta} we show the differential cross sections with respect to the opening angle between the electron-type neutrino and electron, $d\sigma/d\cos\theta_{\nu e}$, for both ES
and CC interactions for three representative neutrino energies: 5, 10, and 20~MeV.
For both ES and CC cross sections, we calculate the angular and electron energy dependence using the MARLEY v1.2.0 neutrino event generator
\cite{marleyCPC,marleyPRC,MARLEYv1.2.0}. We simulate $10^7$ ES scattering events and $6\times 10^6$ CC scattering events with neutrino energies between $0$ and $50$~MeV using MARLEY, and use the resulting binned distributions to generate interpolated functions for scattering angles and electron energies as a function of neutrino energy. These interpolated functions are used throughout this work.
The CC cross sections for $\mathcal{O}(10)$~MeV-scale
$\nu_e-{^{40}\mathrm{Ar}}$ scattering depend non-trivially on nuclear
physics.
In particular, the angular distribution for these interactions is uncertain and depends on nuclear state transition probabilities that are not completely understood \cite{VanDessel:2019obk}. Future measurements of the charged-current differential cross sections at these energies will be crucial to a complete understanding of the signal and backgrounds for the search we describe here, and other measurements in this energy range.

While the final state electron from
ES is highly peaked along the neutrino direction, the CC electron distribution
according to the MARLEY model is nearly flat. The cross-sections for $\bar{\nu}_e$ are $\sim 40\%$ of the $\nu_e$ values, while the $\nu_\mu$ and $\nu_\tau$ cross sections (and their antiparticles) are $\sim 15\%$ of the electron-type. In this work, we will neglect the second-- and third-generation scattering.

In order to calculate limits using directional information, we must calculate the scattering angle $\theta_{\nu e}$ of the electron relative to the neutrino direction, and the correlated electron kinetic energy $E_e$. Both of these distributions are dependent on the neutrino energy $E_\nu$. For ES, the relationship between $E_\nu$, $E_e$, and $\theta_{\nu e}$ can be calculated analytically \cite{Giunti:2007ry}:
\begin{equation}
E_e = \frac{2m_eE_\nu^2 \cos^2\theta_{\nu e} }{(m_e+E_\nu)^2-E_\nu^2\cos^2\theta_{\nu e}}. \label{eq:enuee}
\end{equation}
For CC scattering, the recoiling nucleus can exist in a number of excited states. This makes an analytic relation difficult to calculate, and we rely on MARLEY to numerically simulate the double-differential distributions $d^2 \sigma/d\cos\theta_{\nu e}dE_e$.

\begin{figure}[ht]
\includegraphics[width=0.9\columnwidth]{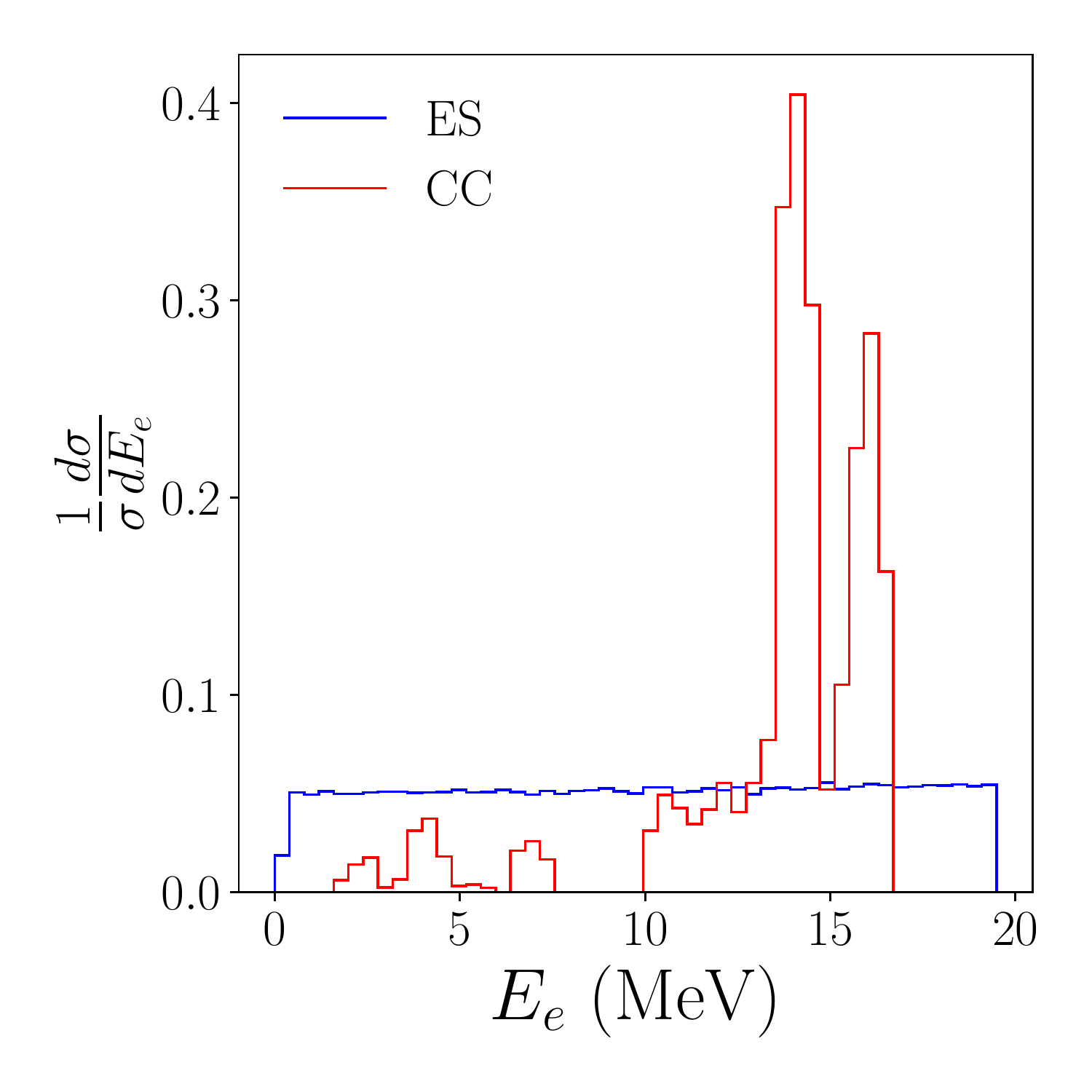} 
\caption{Differential cross section for 20~MeV neutrino scattering via electron scattering (ES, blue) and charged current neutrino-argon scattering (CC, red), calculated through Monte Carlo simulation using MARLEY. The complex scattering structure of CC is due to excited nuclear states within the argon target.
 \label{fig:kinematics}}
\end{figure}
In Figure~\ref{fig:kinematics} we show the distribution of electron kinetic energies $E_e$ in the two scattering modes for incoming neutrino energies $E_\nu = 20$~MeV, again obtained using MARLEY simulation. As can be seen from the combination of Figures~\ref{fig:dsigmadtheta} and \ref{fig:kinematics}, ES scattering results in electrons which are forward-peaked, but with a relatively wide distribution of $E_e$ (compared to $E_\nu$). CC scattering, in contrast, results in scattering which is nearly isotropic, but with electron energies more narrowly distributed and typically closer to $E_\nu$.

Experimental reconstruction of the electron's direction of travel is complicated by a number of effects,
including multiple Coulomb scattering, the angular resolution associated with
the spatial granularity of the readout plane, and the ability of event reconstruction algorithms to
resolve the track direction --- in particular, a possible $180^\circ$
forward-backward ambiguity. We quantify the angular resolution with an overall energy-dependent resolution function $\sigma_{\rm res}(E_e)$.
The resolution parameter can be thought of as the width of the smeared distribution of measured electron directions as compared to the truth.
Studies of supernova event pointing in DUNE reveal the significant challenges in this
energy regime: conventional track reconstruction approaches suffer substantially from
the inherent $180^\circ$ ambiguity in the orientation of an isolated track, leading to
an angular resolution of approximately $140^\circ$ that can be improved to $30^\circ$ at
50~MeV by accounting for associated Bremsstrahlung activity~\cite{DUNE:2020ypp}. The impact
of this improvement declines steadily with energy however, as Bremsstrahlung becomes
less frequent. Resolving this ambiguity alone reduces the angular resolution to $40^\circ$ near threshold,
with otherwise conventional reconstruction tools~\cite{DUNE:2020ypp}.

Several handles are available to further improve reconstruction performance and thus
angular resolution. In addition to Bremsstrahlung activity, the topology of tracks
contains considerable directional information. At the energies we consider ($5-50$~MeV),
electrons will travel at least a few centimeters, crossing several readout channels. The
high ionization density associated with a Bragg peak may be used to help identify
a start and end point, along with the track curvature due to Coulomb scattering, an
effect which becomes larger as the electron loses energy. Finally, the initial segment
of the track, which carries the most information about the initial direction, can be
isolated and used to reconstruct the direction. These possibilities become particularly compelling for pixel-based
readout systems, which provide unambiguous 3D spatial information even for MeV-scale
energy deposits. Such techniques have recently been studied in the context of the QPix
pixel-based charge readout system, where simulations of supernova neutrino ES events
were used to demonstrate an algorithm successfully reconstructing 68\% of single ES scatters within $64^\circ$ of the true direction, including the previously noted ambiguities \cite{Q-Pix:2022zjm}. That reference also includes an example illustration of three-dimensional event within the relevant electron energy range.

In our analysis, we compute two limits: an ``all-sky'' sensitivity which uses no angular information, and a sensitivity assuming a flat $30^\circ$ angular resolution for all electron energies and angles as a bounding case. The latter is chosen to be consistent with the best performance demonstrated in LArTPCs at higher energies, and on par with the resolutions achievable with large-scale water Cherenkov neutrino observatories \cite{SNO:2006odc, Super-Kamiokande:2010tar, Bonventre:2018hyd}, in light of the range of LArTPC performance that depends on rapidly-improving technology and analysis approaches.
Limits depending on directional information using the existing angular resolutions will be mildly weaker, most notably at the lowest electron energies, with the all-sky limits corresponding to the limiting case in which no direction information is available.

In principle, even with limited angular information one can reject neutrinos
that are likely to have originated from the Sun (and to a lesser extent
the isotropic cosmic background) in favor of neutrinos from the Galactic
Center, thus improving sensitivity to signals of dark matter annihilation.
Hence, improved angular resolution leads to reduced acceptance for isotropic
backgrounds, and discrimination of CC and ES final states enables an enhancement of the
ES events that carry nontrivial directional information. 

With our benchmark resolution, we obtain the observed distribution of electron directions $\Omega_{\rm obs}= (\theta_{\rm obs},\phi_{\rm obs})$ relative to the neutrino flight path ${\Omega}_\nu = (\theta_\nu,\phi_\nu)$ by convolving the differential cross section (which depends only on the opening angle between ${\Omega}_{\rm obs}$ and ${\Omega}_\nu$) with a Gaussian error function on $\Omega_{\rm obs}$ that has a standard deviation $\sigma_{\rm res}$. Practically, we calculate the distribution
\[
\frac{d^2\sigma}{d\cos\theta_{\rm obs}dE_e}
\]
by randomizing the path of the scattered electron in ES and CC events simulated by MARLEY, then binning and interpolating the results. In the bottom panel of Figure~\ref{fig:dsigmadtheta}, we show the differential cross section $d\sigma/d\cos\theta_{\rm obs}$ assuming the flat $30^\circ$ resolution and integrating over all electron energies exceeding a 5~MeV threshold.

Given a flux $\Phi_\nu$ of neutrinos that depends on angular position $\Omega = (\theta,\phi)$, the recoiling electrons have an observed spectrum with an angular dependence on $\Omega_{\rm obs}$ of
\begin{equation}
\frac{d^2\Phi_e(E_\nu)}{d\Omega_{\rm obs}dE_e} = \int d\Omega \frac{d^2\sigma}{d\cos\theta_{\nu e}dE_e} \frac{d\Phi_\nu(E_\nu)}{d\Omega}, \label{eq:phiconvolved}
\end{equation}
where $\cos\theta_{\nu e}$ is the opening angle between $\Omega$ and $\Omega_{\rm obs}$. 
The resulting observed angular distribution of electrons from neutrino scattering events is then
\begin{equation}
\frac{d^2N_e}{d\Omega_{\rm obs} dE_e} = (N_{\rm target} \times T) \times \epsilon(E_\nu) \times \frac{d^2\Phi_e(E_\nu)}{d\Omega_{\rm obs}dE_e},  \label{eq:Nconvolved}
\end{equation}
where $N_{\rm target} \times T$ is the exposure of the detector, and $\epsilon(E_\nu)$ is the energy-dependent detector efficiency. We assume an exposure of 40~kton$\times$year. Based on Ref.~\cite{DUNE:2020ypp}, we assume a detector efficiency of $100\%$ for $E_e >5$~MeV and zero below this threshold. This implies a minimum neutrino energy of $\sim 6$~MeV. 

For computational simplicity, for the remainder of this work we will use electron energy bins of width 1~MeV, larger than the expected energy resolution for a LArTPC detector \cite{Friedland:2018vry}. Finer binning may result in improvements of our predicted limits, at the cost of increased computational time.

\subsection{CC/ES Discrimination}

Finally, we consider the impact of distinguishing the ES scattering events from the CC. The reduced sensitivity of the detector to new physics due to lower overall signal rate can be outweighed by a greater decrease in the background rate once directional information is factored in, as the scattered electrons will be more closely aligned with the original neutrino direction, which can improve sensitivity to a signal from a well-localized source.
Previous work has explored the possibility to discriminate between
CC and ES scatters on an event-by-event basis in large-scale LArTPCs
based on the presence or absence of correlated final state deexcitation
photon activity \cite{Castiglioni:2020tsu}. Based on that work, we
assume a sample of ES events can be obtained with an efficiency
$\epsilon_{\rm ES}$, defined as the fraction of all ES events
that would pass the discrimination cuts, and with a CC contamination
given by a constant efficiency $\epsilon_{\rm CC}$ \cite{Castiglioni:2020tsu,Lepetic:2021}:
\begin{equation}
\epsilon_{\rm ES} = e^{-0.0464\times (E_\nu/{\rm MeV})},\quad \epsilon_{\rm CC} = 0.0072. \label{eq:escc_efficiency}
\end{equation}

\section{Dark Matter Neutrino Signals \label{sec:signals}}
Having calculated the angular dependence of the observed electrons inside the detector as related to the neutrino signal, we now apply these results to signal of dark matter annihilation in the Galactic Center. Such neutrinos would be highly localized towards the Galactic Center. 
If some of this directionality is imprinted into experimental measurement, it can be used to enhance signal over background, especially the background from highly-localized Solar neutrinos.

The differential flux of neutrinos from annihilating dark matter of mass $m_\chi$ and velocity-averaged cross section $\langle \sigma v\rangle$ as a function of energy $E$ is given by
\begin{equation}
\frac{d\Phi_\chi(E_\nu)}{d\Omega} = \frac{\langle \sigma v \rangle}{8 \pi m_\chi^2} \frac{dN}{dE_\nu} \int_{\rm l.o.s.} d\ell \rho_\chi^2(\ell,\Omega). \label{eq:DMnuflux}
\end{equation} 
In this section, we assume $\langle \sigma v \rangle$ includes annihilation into all three generations of neutrinos, resulting in equal fluxes of $\nu_e$, $\nu_\mu$, and $\nu_\tau$ (and their antimatter counterparts). As previously discussed, we assume our signal events originate only from $\nu_e$ and to a lesser extent $\bar{\nu}_e$, ignoring a small admixture from the second and third generations. This results in slightly conservative limits.
In Eq.~\eqref{eq:DMnuflux}, $dN/dE_\nu$ is the spectrum of neutrinos emitted by a single dark matter annihilation, which we assume to be monoenergetic, $dN/dE_\nu = 2\delta(E_\nu-m_\chi)$.
The last term in Eq.~\eqref{eq:DMnuflux} is the differential $J$-factor,
$dJ/d\Omega$, a line-of-sight integral of the dark matter density containing all the information about the angular dependence of the signal. The details of the $J$ factor calculation are provided in Appendix~\ref{app:jfactor}.

\begin{figure*}[htbp]
\includegraphics[width=1.8\columnwidth]{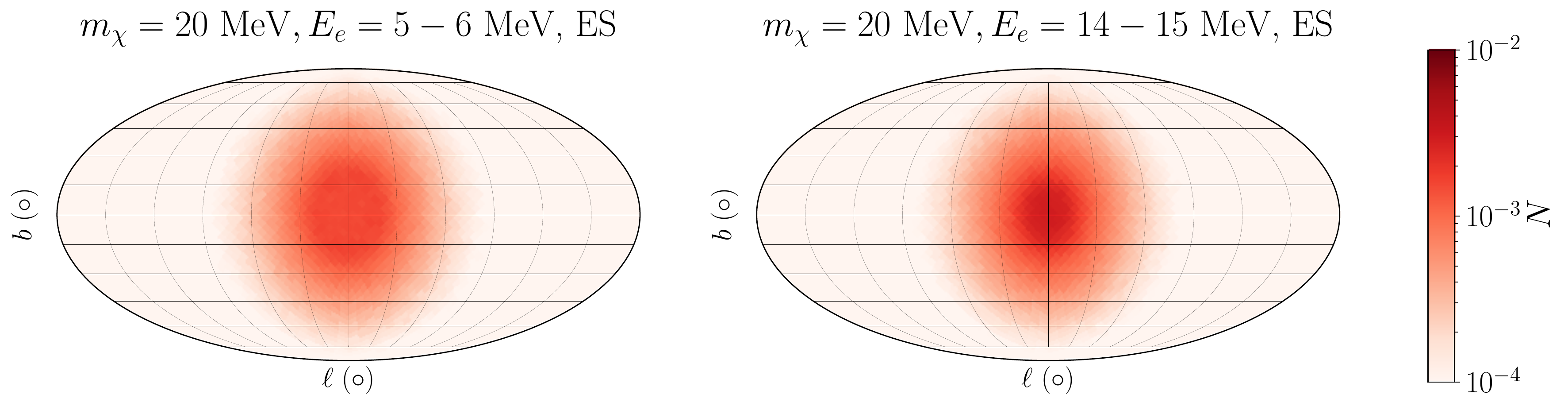}\\
\includegraphics[width=1.8\columnwidth]{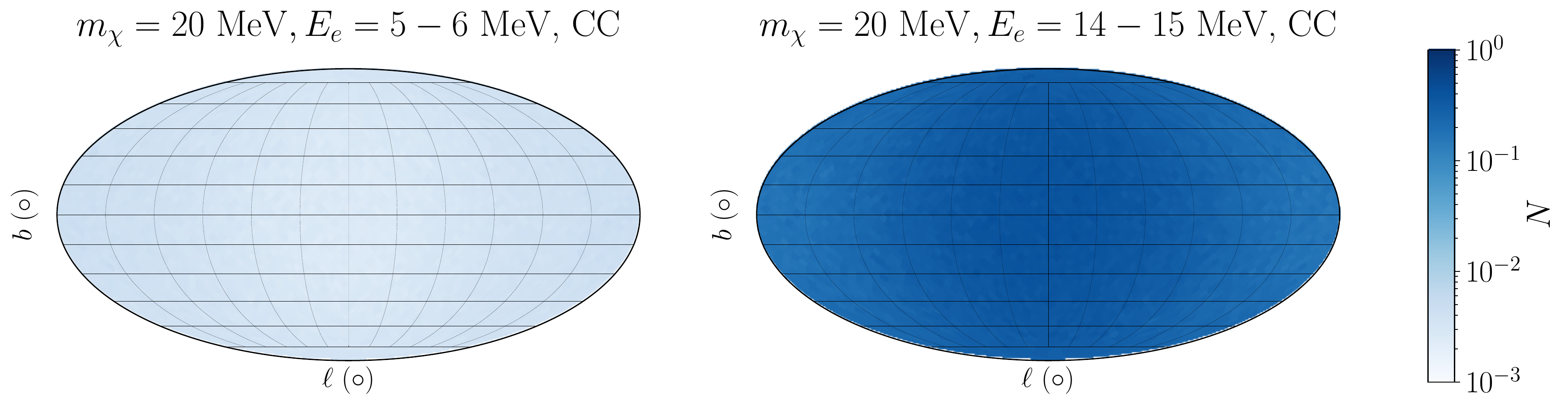}
\caption{Distribution on the sky (in Galactic coordinates centered on the Galactic Center) of the expected number of signal electrons $N$ in angular bins of solid angle $d\Omega = 4\times 10^{-4}$~sr, assuming $m_\chi = 20$~MeV, and $\langle \sigma v\rangle = 10^{-22}$~cm$^3$/s, a LArTPC detector with 40~kton$\times$year of exposure, an angular resolution of $30^\circ$ for electrons. Distributions are shown for ES scattering (upper row) and CC scattering (lower row), and electron energies of $5-6$~MeV (left column) and $14-15$~MeV (right column).
\label{fig:dNdOmega}}
\end{figure*}

\begin{figure}[ht]
\includegraphics[width=0.9\columnwidth]{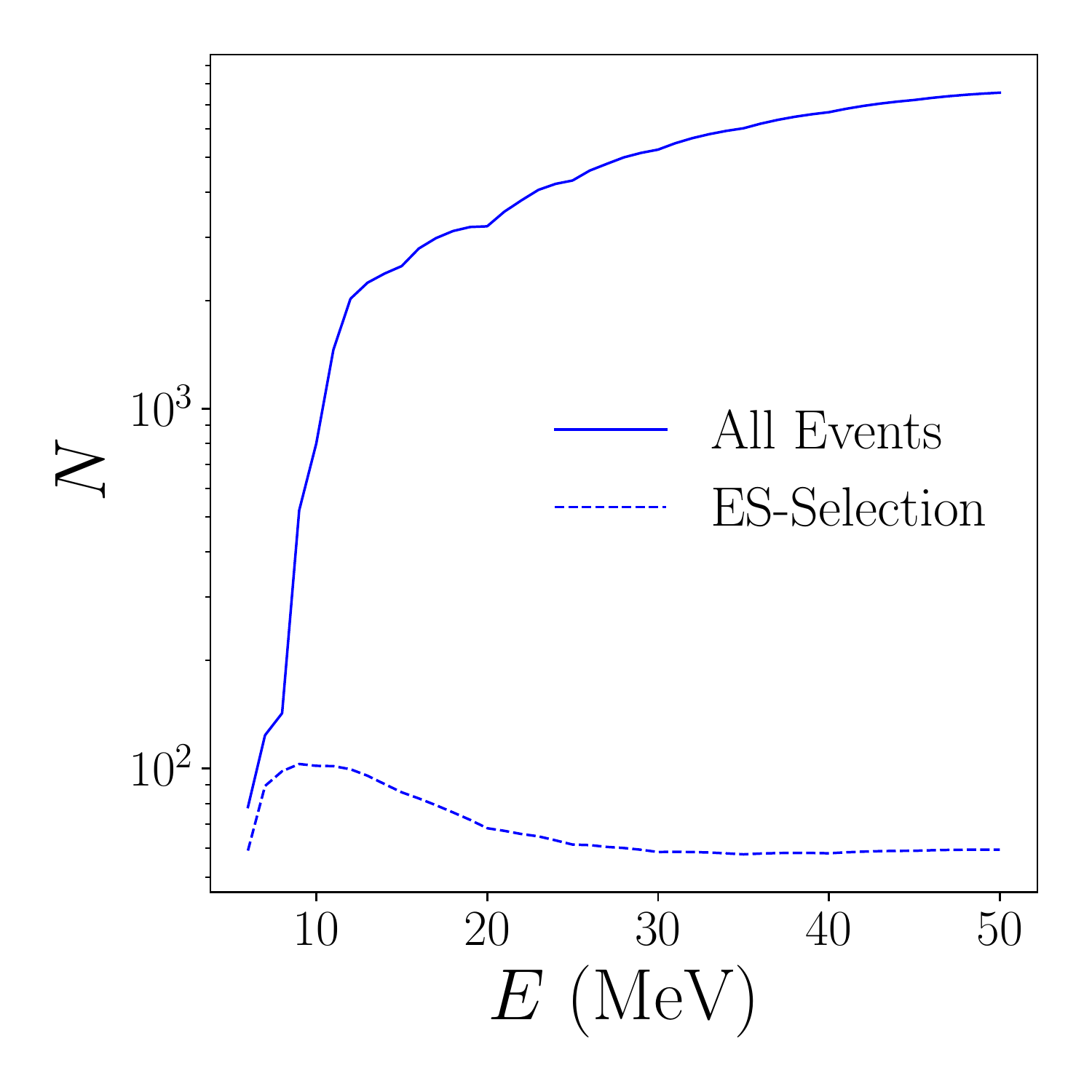}
\caption{Total number of dark matter signal events integrated over the full sky as a function of neutrino energy $E_\nu$, as seen in 40~kton$\times$year of exposure of a LArTPC detector. We assume the best-fit NFW profile of Ref.~\cite{Fornasa:2013iaa} and an annihilation cross section of $\langle \sigma v\rangle = 3\times 10^{-22}$~cm$^3$/s. The solid line shows the prediction for all scattering events, while the dotted line shows those events selected to be ES-like, assuming a selection efficiency of Eq.~\eqref{eq:escc_efficiency} on top of the overall efficiency.
\label{fig:DMnumber}}
\end{figure}

The dark matter-induced neutrino flux must then be convolved with the scattering cross section to give the angular dependence of the visible electron counts, as per Eq.~\eqref{eq:Nconvolved}. We show this angular distribution across the sky in Figure~\ref{fig:dNdOmega}. As can be seen, the CC scattering erases nearly all of the angular dependence of the initial neutrino flux. ES events on the other hand still peak towards the Galactic Center, due to the closer correlation between the neutrino and recoiling electron path. In Figure~\ref{fig:DMnumber}, we show the number of expected signal events detected by a DUNE-like LArTPC detector after integration over the full sky, assuming an annihilation cross section of $\langle \sigma v\rangle = 3\times 10^{-23}$~cm$^3$/s into all three generations of neutrinos at the Galactic Center (of which only the electron-type are detected) and a 40~kton$\times$year exposure.

\section{Backgrounds \label{sec:backgrounds}}

In the previous section, we have determined the expected rate of neutrino events in a LArTPC detector, as a function of the apparent direction of travel of the electrons within the detector. As seen in the previous section, the electron direction will still be correlated with the location of the Galactic Center, especially for ES events. We now turn to estimating background events, and determine the angular distributions of the background electrons.
We consider five primary sources of background in the detector:
scattering of $^8$B and $hep$ solar neutrinos, atmospheric neutrino
interactions, decays of radio-isotopes produced via spallation by
cosmic ray muons, and the yet-undetected diffuse supernova neutrino
background (DSNB).

For the solar neutrinos, spallation, and DSNB,
we adopt the model of Zhu, Beacom, and Li \cite{Zhu:2018rwc},
developed in
the context of DUNE solar neutrino sensitivity \cite{Capozzi:2018dat},
within the same
signal energy range. In particular, we assume the same observed
electron spectra as obtained in that work, including a suite of
spallation-induced background reduction cuts, and scale these to
our nominal detector exposure. We note that these electron energy
distributions include a 7\% energy resolution, while the
effect of final state electron energy resolution is not included
in our signal model; the effect of this approximation on the
sensitivity is minimal due the relatively coarse 1~MeV energy bins
used in our analysis.

Atmospheric neutrino interactions dominate the background in the
energy region above the solar neutrino endpoint. To approximate this
background, we consider the model of Cocco et al. \cite{Cocco:2004ac}.
That work
presents a detailed analysis of the DSNB sensitivity of the ICARUS
LArTPC detector \cite{ICARUS:2004wqc} sited underground at
the INFN Gran Sasso Laboratory (LNGS).
This includes a prediction of the atmospheric $\nu_e$
flux and the electron energy spectra for backgrounds based on a
detailed FLUKA \cite{Ferrari:2005zk,Bohlen:2014buj} Monte Carlo
simulation. The atmospheric
neutrino flux will depend to some extent on the detector location,
in particular the geomagnetic latitude \cite{Gaisser:2002jj}
and the time of
exposure relative to the solar cycle, leading to variations on the order of $\lesssim25\%$. Meanwhile, the observed electron
spectrum is also dependent on detector effects, reconstruction, and analysis strategies for background mitigation, among other factors, which is likely a larger effect. In view of this, we treat
the results for ICARUS at LNGS presented in Ref.~\cite{Cocco:2004ac} as representative of the
relevant atmospheric neutrino backgrounds for a large scale, deep-underground LArTPC detector centrally located in the Northern middle latitudes.

\begin{figure}[th]
\includegraphics[width=0.9\columnwidth]{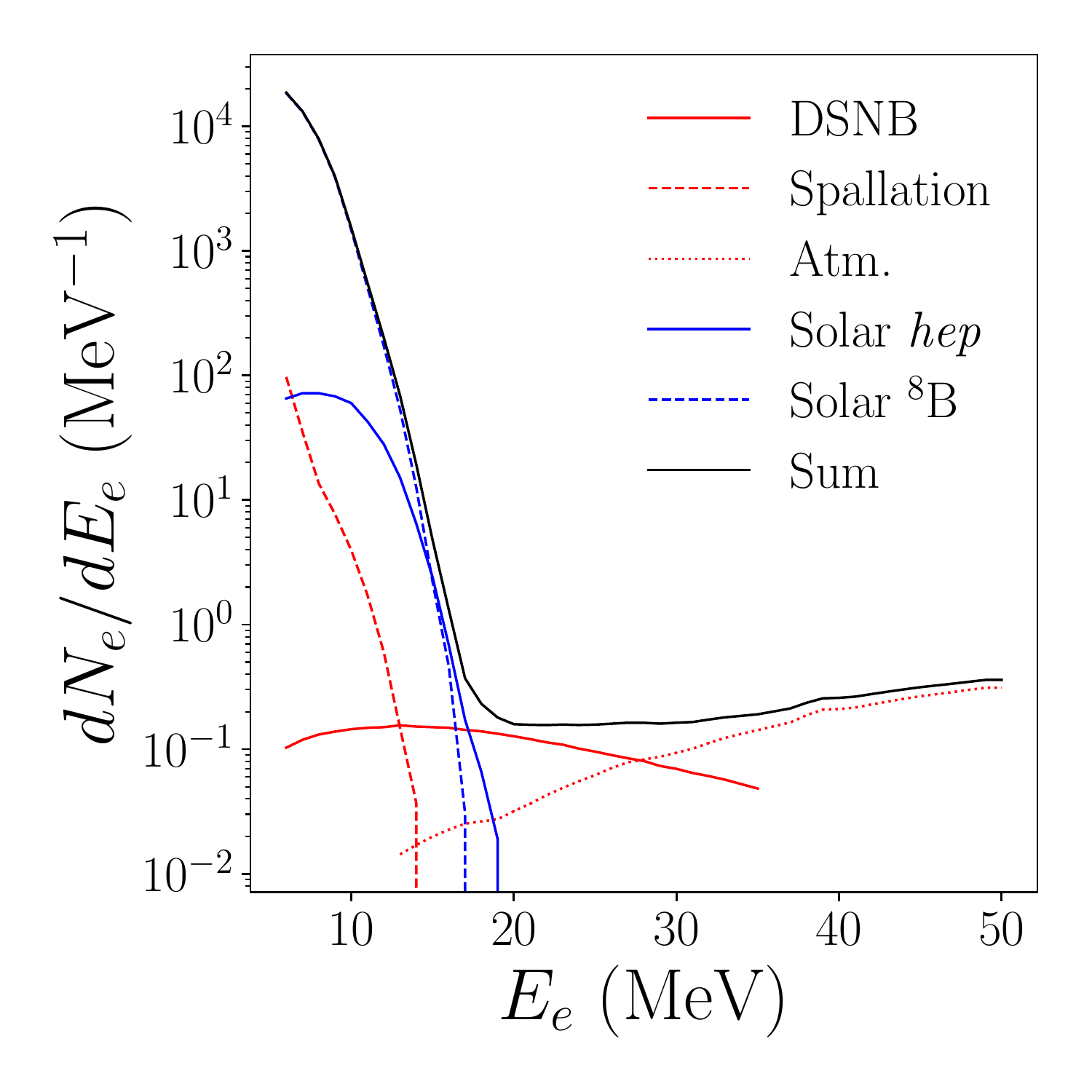}\\
\caption{
Differential rate of background electron events in a LArTPC detector as a function of electron energy (per 40 kton$\times$yr exposure), based on Refs.~\cite{Zhu:2018rwc, Capozzi:2018dat, Cocco:2004ac} (see text for details). The diffuse supernova neutrino background (DSNB), spallation, and atmospheric backgrounds are isotropic, while the two Solar backgrounds come from the direction of the Sun.
\label{fig:backgrounds}}
\end{figure}

Of the background categories we include, atmospheric neutrinos, spallation-induced decays, and the DSNB are essentially isotropic, and yield isotropic scattered electron distributions. The Solar neutrinos emanate from the core of the Sun, which we take to be a point source on the sky. In Figure~\ref{fig:backgrounds}, we show the expected differential rate of observed background events with our assumed exposure.
As can be seen, below 20~MeV, the Solar neutrinos dominate, suggesting that angular information might be especially useful to reject background in this regime. 

The angular dependence of the observed electrons relative to the Sun can be calculated through Eq.~\eqref{eq:phiconvolved}, taking the differential flux of neutrinos to be a delta function at the Sun's location and assuming each background has a source spectrum of {\it neutrino} energies, which are taken from Ref. \cite{Billard:2013qya}.
The electrons due to spallation-induced activity have no relevant parent neutrino spectrum, and we assume this background is isotropic.
\begin{figure}[htbp]
\includegraphics[width=0.9\columnwidth]{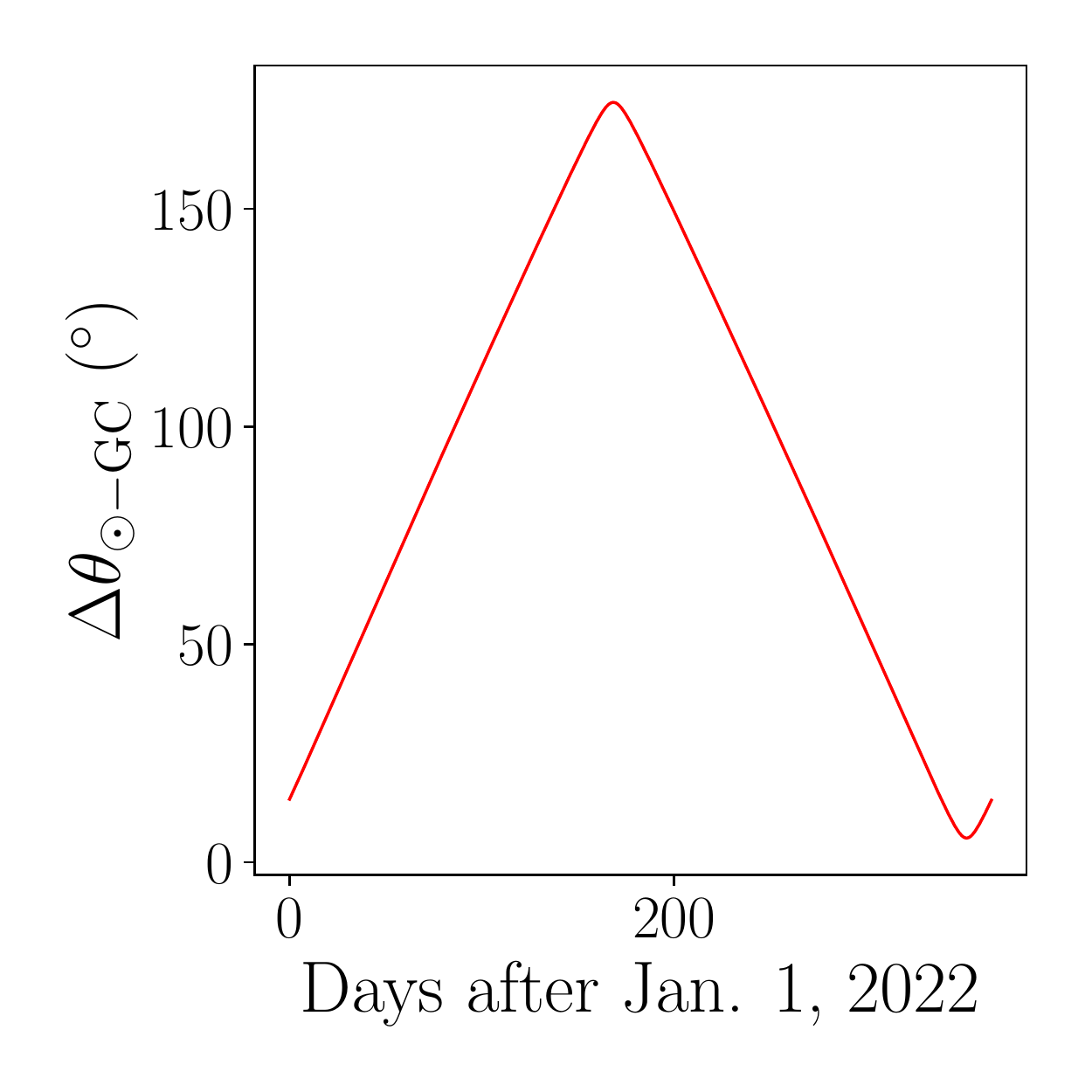}
\caption{Opening angle between the Sun and the Galactic Center as seen from Earth as a function of the date after January 1, 2022 (using the Solar position from \textsc{AstroPy} \cite{astropy:2013,astropy:2018}). \label{fig:solar_angle}}
\end{figure} 

\begin{figure*}[htbp]
\includegraphics[width=1.8\columnwidth]{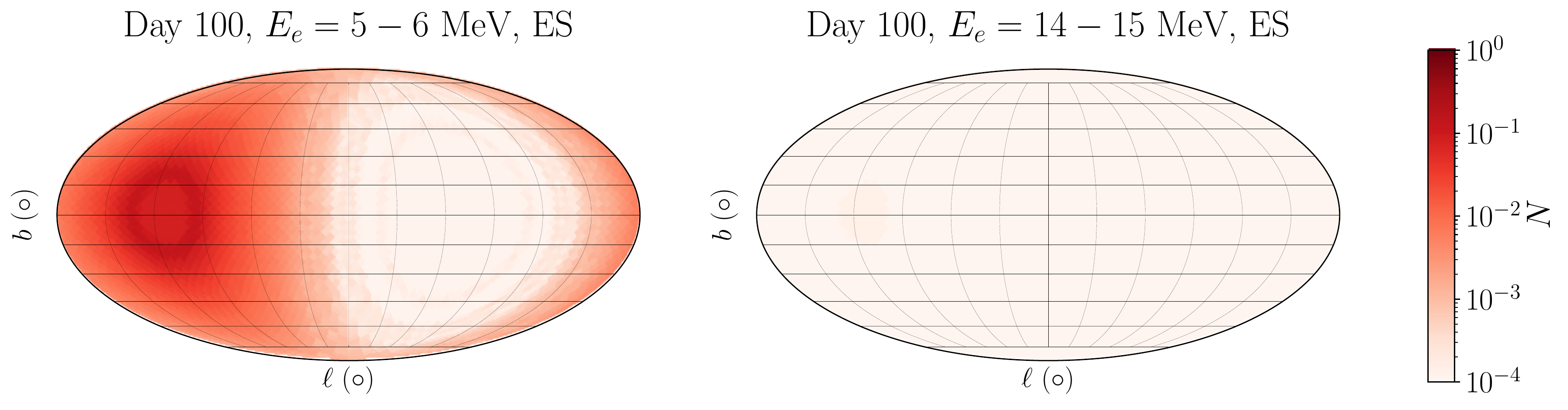}
\includegraphics[width=1.8\columnwidth]{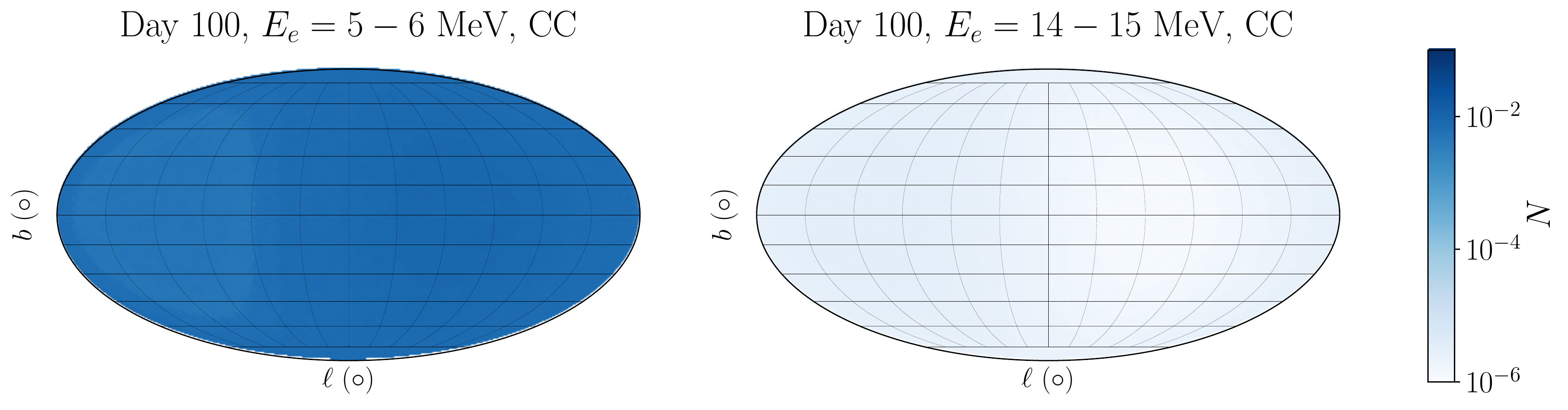}
\caption{Distribution on the sky (in Galactic coordinates centered on the Galactic Center) of the expected number of background electrons $N$ on April 10$^{\rm th}$ (Day 100 of Figure~\ref{fig:solar_angle}) in angular bins of solid angle $d\Omega = 4\times 10^{-4}$~sr, a LArTPC detector with 40~kton$\times$year of exposure, and an angular resolution of $30^\circ$ for electrons. Distributions are shown for ES-scattering (upper row) and CC-scattering (lower row), and electron energies of $5-6$~MeV (left column) and $14-15$~MeV (right column).
\label{fig:background_days}}
\end{figure*}

The Sun moves relative to the Galactic Center over the course of a year, as shown in Figure~\ref{fig:solar_angle}. Therefore, for any location on the sky (measured against the fixed stars and thus the Galactic Center), the relative rate of signal and background events will be continually changing throughout the year. In Figure~\ref{fig:background_days} we show the electron background rate for one day in the year (chosen as April 10$^{\rm th}$, 2022, or Day 100), demonstrating the angular dependence of the Solar backgrounds in both ES and CC channels.
In the next section, we describe our approach to maximize the projected sensitivity, using day-by-day predictions for the signal and background differential rates to fold in directional information to our limits.

\section{Model Independent Limits \label{sec:limits}}

If directional information were not available, we can set predicted limits on the annihilation cross section of dark matter into neutrinos in the Galactic Center by comparing the predicted signal rate to the number of background events in each 1~MeV bin of electron energy, from our threshold of 5~MeV up to the maximum possible $E_e\sim m_\chi$. This ``all-sky'' limit results in strictly weaker limits if ES-like events are selected, as both signal and background event rates would be reduced by the same factor.

By adding the apparent direction of the scattered electron, additional information can be used to set the limit: the probability of any individual event being the result of signal from the Galactic Center versus originating from the isotropic plus Solar backgrounds. In this case, the correlation between the neutrino and electron directions in ES events can result in a significantly improved bound if ES-like events can be isolated within the LArTPC.

To incorporate both the event rate and the probability given the direction of the scattered electron in our statistical treatment, we use the CL$_{\rm s}$ method to set limits. We account for the changing relative location of the Sun by considering data acquisition day-by-day over the year. For each day, we calculate the Solar location relative to the Galactic Center, and generate a random sample of background and signal events across the sky, using the differential distributions calculated in Sections~\ref{sec:signals} and \ref{sec:backgrounds}. 

Specifically, to generate the differential distributions, we divide the sky into 3072 equal bins using \textsc{HealPy}\cite{2005ApJ...622..759G,Zonca2019}, use bins of width 1~MeV in $E_e$ from 5~MeV up to the assumed value for $m_\chi$, calculate an expected signal and background rate in each angular and energy bin for each day, and draw an expected number of events in each bin from a Poisson distribution with the bin-dependent expected event rate.  The overall normalization of the signal rate depends on the assumed value of $\langle \sigma v \rangle$. The angular and energy bin sizes are chosen to be sufficiently small while remaining computationally tractable; smaller bin sizes would increase the power of each neutrino event to statistically discriminate between signal and background, at the cost of increased analysis time.

We then define a log-likelihood ratio for the observed number of events in each bin $i$ that corresponds to observed electron direction $\Omega_i$ and energy $E_\alpha$ on day $t$ as
\begin{eqnarray}
& L(\langle \sigma v\rangle,\Omega_i,E_\alpha,t) = &  \\
& \ln \Big( f\big[ n_{\rm obs}(\Omega_i,E_\alpha,t) \big| n_s(\Omega_i,E_\alpha,t,\langle \sigma v\rangle)+n_b(\Omega_i,E_\alpha,t)\big]\Big) & \nonumber \\
& -\ln \Big( f\big[ n_{\rm obs}(\Omega_i,E_\alpha,t) \big| n_b(\Omega_i,E_\alpha,t) \big]\Big), & \nonumber
\end{eqnarray}
where $f(\lambda |n)$ is the Poisson distribution for $\lambda$ observed events with an expectation of $n$ events, $n_s(\Omega_i,E_\alpha,t,\langle \sigma v\rangle)$ is the expected number of signal events in bin indexed by angular location $\Omega_i$ and electron energy $E_\alpha$ on day $t$ assuming dark matter annihilation cross section $\langle \sigma v\rangle$, and $n_b(\Omega_i,E_\alpha,t)$ is the expected number of background events in that bin over that day. In the ``all-sky'' analysis, we collect the simulated events in a single angular bin.

We then construct the signal and background distributions of $\sum_{i,t} L(\langle \sigma v\rangle,\Omega_i,E_\alpha,t_i)$, where the sum runs over all angular and energy bins and every day of the data-taking (assumed to be one full year). First, we generate 1,000 iterations of the events expected from 40~kton$\times$year of exposure, assuming the presence of signal with cross section $\langle \sigma v\rangle$. From this, we can construct the probability distribution of the log-likelihood in the presence of signal, $P_{\rm sb}\left(\sum_{i,t} L(\langle \sigma v\rangle,\Omega_i,E_\alpha,t)\right)$ through a histogram of the summed log-likelihood. Next, events are generated assuming background only, constructing $P_{\rm b}\left(\sum_{i,t} L(\langle \sigma v\rangle,\Omega_i,E_\alpha,t)\right)$ out of the histogram of 1,000 sets of mock background-only observations. The resulting limits assuming a smaller detector volume and a longer period of exposure would be very similar, but computationally more expensive.

The CL$_{\rm sb}(\langle \sigma v\rangle)$ and CL$_{\rm b}(\langle \sigma v\rangle)$ parameters for a set of observed neutrino events $\{\Omega_i,E_\alpha,t\}$ --- assuming a signal event rate set by an annihilation cross section $\langle \sigma v\rangle$ --- are then defined as
\begin{eqnarray}
{\rm CL}_{\rm sb} & = & \int_{x>\sum_{i,t} L(\langle \sigma v\rangle,\Omega_i,E_\alpha,t)}^\infty dx P_{\rm sb}\left(x \right) \label{eq:clsb_int} \\
{\rm CL}_{\rm b} & = & \int_{x>\sum_{i,t} L(\langle \sigma v\rangle,\Omega_i,E_\alpha,t)}^\infty dx P_{\rm b}\left(x \right). \label{eq:clb_int}
\end{eqnarray}
That is, CL$_{\rm sb}$ and CL$_{\rm b}$ are the probabilities of seeing a set of events more signal-like than what was actually observed, assuming the presence (for CL$_{\rm sb}$) or absence (for CL$_{\rm b}$) of signal. The exclusion limit CL$_{\rm s}$ for a given cross section is then the ratio
\begin{equation}
{\rm CL}_{\rm s}\left( \{\Omega_i,E_\alpha,t\},\langle \sigma v\rangle\right) = \frac{{\rm CL}_{\rm sb}\left( \{\Omega_i,E_\alpha,t\},\langle \sigma v\rangle\right) }{{\rm CL}_{\rm b}\left( \{\Omega_i,E_\alpha,t\},\langle \sigma v\rangle\right) }
\end{equation}
To obtain projected limits, we set the observed events to be the expected value for background only (that is, CL$_{\rm sb} = 0.5$) and calculate a limit for the cross section when CL$_{\rm s} = 0.05$ (i.e., 95\% exclusion).

\begin{figure}[ht]
\includegraphics[width=0.9\columnwidth]{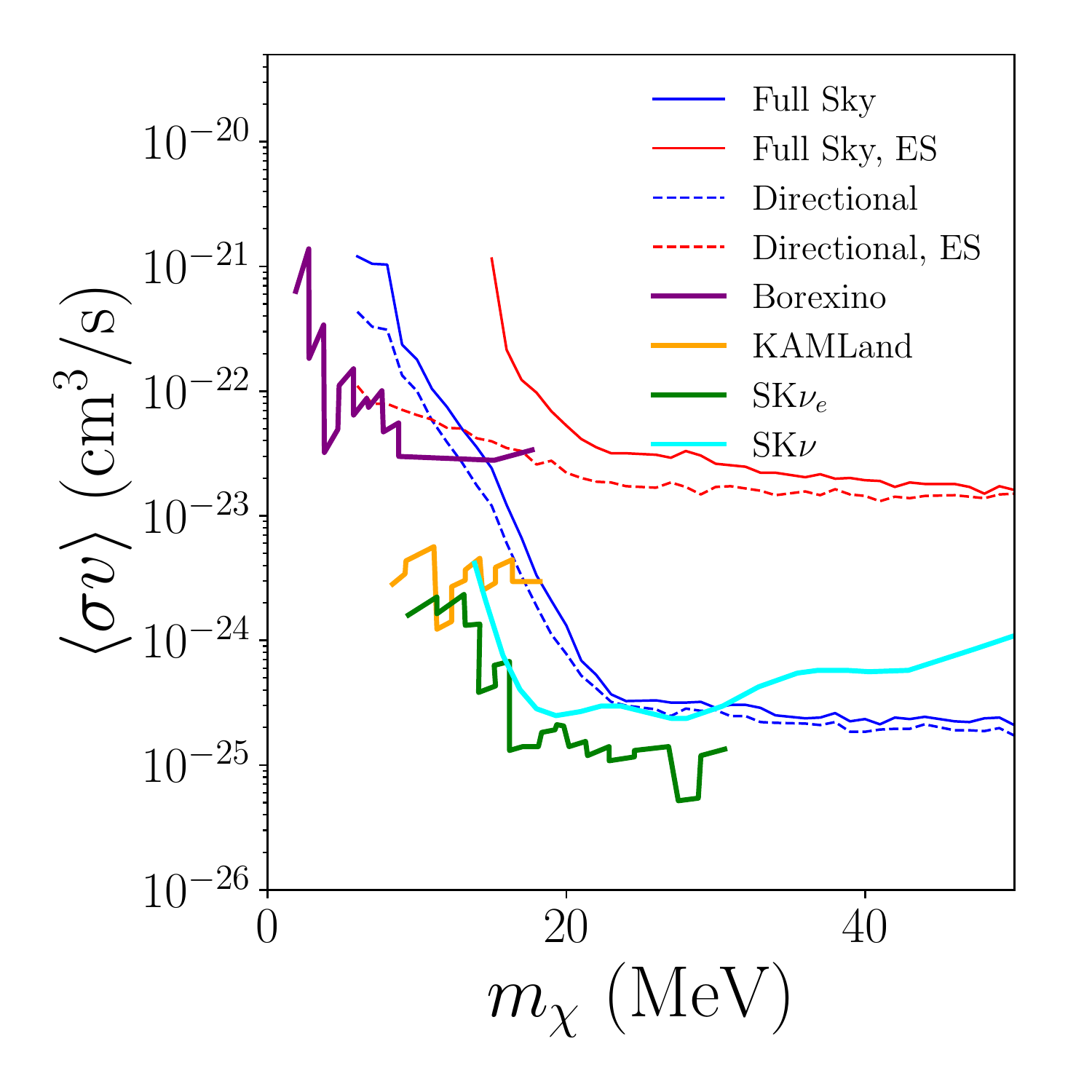}
\caption{Projected 95\% CL upper limits on dark matter annihilation cross section $\langle \sigma v\rangle$ into all three flavors of neutrinos as a function of dark matter mass, assuming 40 kton$\times$year exposure. All-sky events (no angular information) are shown in solid lines,
and assuming $\sigma_{\rm res} = 30^\circ$ in dashed. Limits considering both ES and CC events are in blue, the ES-selected subsample limits are in red. Existing limits using Borexino \cite{2011PhLB..696..191B} (purple), KamLAND \cite{KamLAND:2011bnd,KamLAND:2021gvi} (orange), Super-Kamiokande 
\cite{Super-Kamiokande:2015qek} $\bar{\nu}_e$ (green), and Super-Kamiokande diffuse supernovae flux search \cite{Olivares-DelCampo:2017feq} (cyan) data as reanalyzed in Ref.~\cite{Arguelles:2019ouk} are shown for comparison. \label{fig:model_indep_sigmav}}
\end{figure}

In Figure~\ref{fig:model_indep_sigmav}, we show the predicted upper limits at 95\% confidence for $\langle \sigma v\rangle$ from 40 kton$\times$year exposure. Limits for two sets of assumptions are shown:
\begin{enumerate} 
\item the ``all-sky'' analysis, which ignores directional information, and
\item limits using directional information, assuming an angular resolution of $30^\circ$ for electrons.
\end{enumerate} 
For both of these options, we consider analyses that use all scattering events (i.e., both CC and ES processes) as well as an analysis after ES selection has been applied. The latter category contains some admixture of CC scattering, through the efficiency factors of Eq.~\eqref{eq:escc_efficiency}.

As expected, we see that without directional information, the ES-selected subsample results in strictly weaker limits. At high dark matter mass (and thus neutrino and electron energies), the cross section for electron scattering is relatively isotropic and directional information does not significantly improve these limits (either with or without the ES selection). Above 30~MeV, above the range of the Super-Kamiokande limits derived using $\bar{\nu}_e$ events \cite{Super-Kamiokande:2015qek}, a DUNE-like LArTPC detector could exceed all existing limits, using all events and regardless of directional information. It is likely these relatively flat limits would extend above the 50~MeV maximum mass we consider in this work.\footnote{Expected limits for higher dark matter masses were not considered as the computational time over many bins of electron energy became prohibitive given the available resources.}

In the lower energy regime, $m_\chi \lesssim 20$~MeV, where Solar backgrounds dominate and the more-colinear ES events are relevant, directional information can significantly improve the limits, up to nearly an order of magnitude (relative to the all-sky, all-event analysis) at the lowest energies, when employing a selection to enhance the purity of ES-like events.
 
\section{Model Dependent Interpretation \label{sec:interpretations}}
As a concrete example of our projected limits, we now apply our model-independent results to the parameter space of a representative well-motivated model of dark matter with strong couplings to neutrinos. We consider a gauged vector portal model  in which the Standard Model is extended to include an anomaly-free $U(1)_{L_{\mu} - L_{\tau}}$ gauge group \cite{Ma:2001md}. The spontaneous breaking of this gauge symmetry leads to a massive gauge boson $Z^{\prime}$ which interacts only with $\mu$, $\tau$, $\nu_{\mu}$, $\nu_{\tau}$, and their corresponding anti-particles.
Further details of the model including the Lagrangian of interactions between Standard Model fermions $f$ and the dark matter $\chi$ are given in Appendix~\ref{app:dmmodel}.

In this model, a mediator $Z^{\prime}$ interacts primarily with $2^{\rm nd}$ and $3^{\rm rd}$ generation Standard Model leptons through a gauge coupling $g_{f}$. However, a non-zero kinetic mixing between $Z^{\prime}$ and the neutral Standard Model gauge bosons is induced at the one-loop level \cite{Escudero:2019gzq}.
All the other Standard Model fermions couple to $Z^{\prime}$ through kinetic mixing. As this interaction is subdominant to the interaction with second and third generation leptons, we do not consider it further in this study.

If dark matter $\chi$ is charged under this extra group, then the vector boson can mediate interactions between $\chi$ and the Standard Model. We assume that the dark matter is vector-like under this new gauge group, hence its charge can vary away from unity. Following the convention in Ref.~\cite{Kahn:2018cqs}, we choose three benchmark dark matter models, wherein the dark matter is treated as a Dirac fermion, Majorana fermion, or complex scalar, as defined in Eq.~\eqref{eq:dm_models}.

If there is no particle asymmetry in the dark sector, then the relic abundance in the early Universe is given by the thermal freeze-out of dark matter annihilating to Standard Model particles through the $L_{\mu} - L_{\tau}$ gauge boson $Z^{\prime}$.
The main annihilation processes depend on the mass hierarchy in the dark sector. If $m_{\chi} > m_{Z^{\prime}}$, the dominant process is secluded annihilation of dark matter into the $Z'$, $\chi \chi \rightarrow Z^{\prime} Z^{\prime}$, followed by decay into Standard Model fermions $Z^{\prime} \rightarrow \bar{f} f$.
On the other hand, if $m_{\chi} < m_{Z^{\prime}}$, then the dominant process is the direct $s$-channel annihilation of dark matter through the exchange of a $Z^{\prime}$, i.e., $\chi \chi \rightarrow \mu^{+} \mu^{-}, \tau^{+} \tau^{-}, \nu_{i} \bar{\nu}_{i}$, with $i = {\mu, \tau}$.
The secluded annihilation case could have very interesting phenomenology \cite{Asai:2020qlp}, however it is beyond the scope of this work. 
In this work, as an illustrative example we will focus on the case with $m_{\chi} ~\textless~ m_{Z^{\prime}}$.
\begin{figure*}[ht]
\centering
\includegraphics[width=1.0\columnwidth]{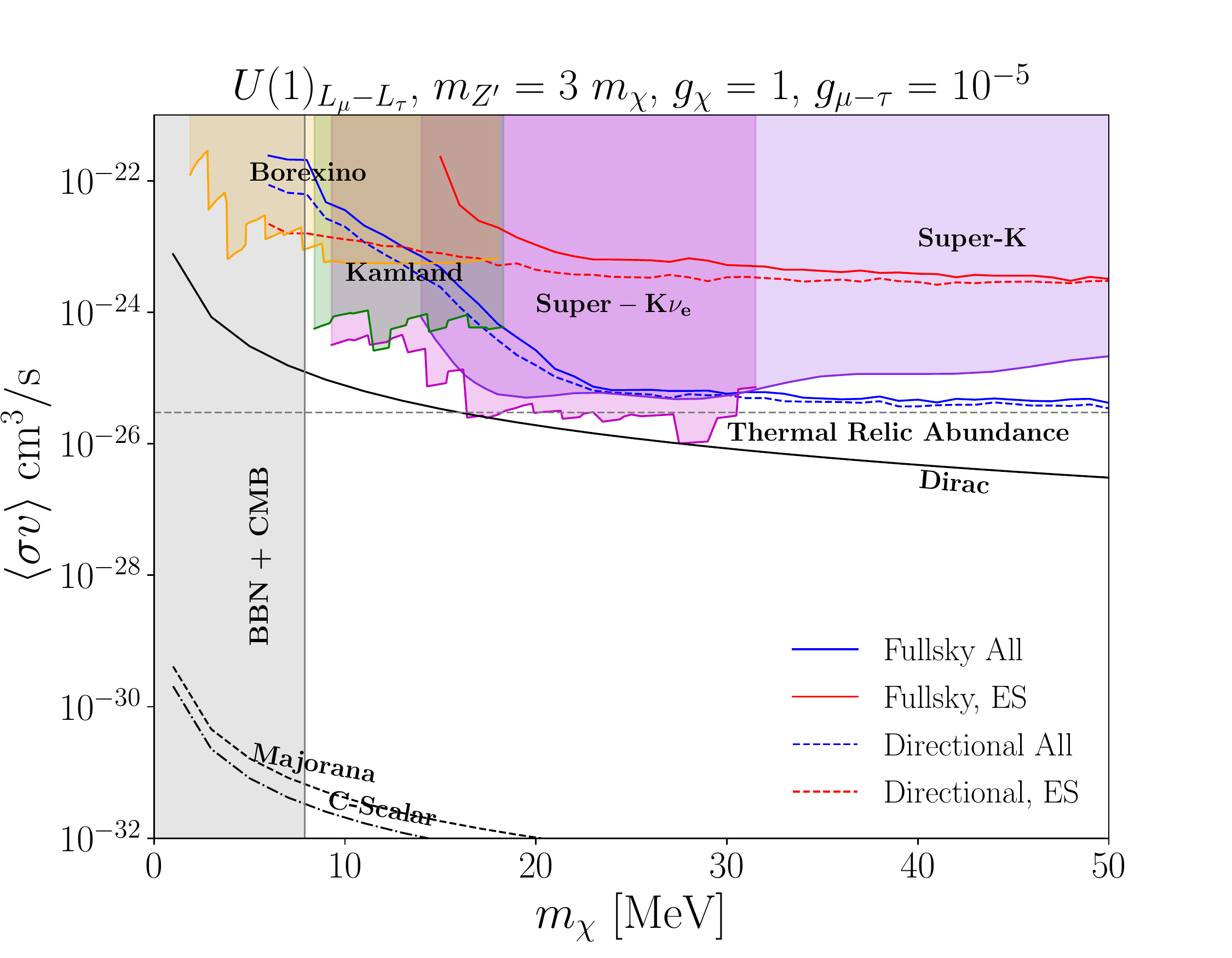}
\includegraphics[width=1.0\columnwidth]{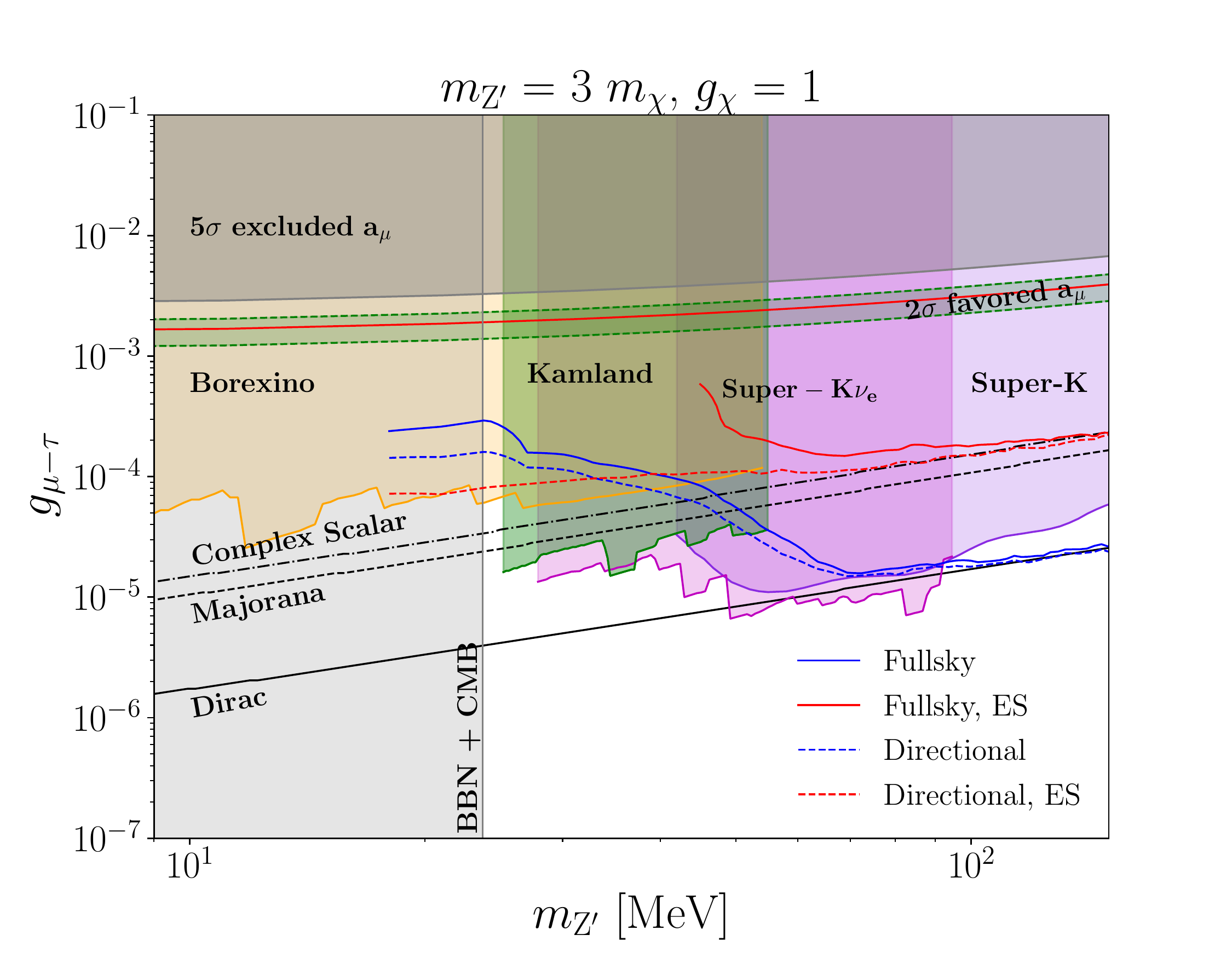}
\caption{Constraints on the gauged $U(1)_{L_{\mu}-L_{\tau}}$ vector portal model parameter space. 
{\it left}: Upper limits on the thermally averaged annihilation cross-section for present time dark matter annihilation into neutrinos. The dashed grey horizontal line represents the thermal relic dark matter that decoupled in the early universe. The darker grey solid, dashed and dot-dashed lines represent the annihilation cross-sections for the Dirac, Majorana and complex scalar models respectively.
{\it right}: Parameter space for dark matter interacting with Standard Model through new vector $Z^{\prime}$. We choose benchmark parameters, $m_{Z^{\prime}} = 3m_{\chi}$ and dark sector gauge coupling $g_{\chi} = 1$. The grey shaded region is excluded by muon $g$-2 data at $5\sigma$ level, while the area inside the green dashed band is allowed with the central value given by the red solid line.
Existing limits using Borexino \cite{2011PhLB..696..191B}, KamLAND \cite{KamLAND:2011bnd,KamLAND:2021gvi}, Super-Kamiokande
\cite{Super-Kamiokande:2015qek} $\bar{\nu}_e$, and Super-Kamiokande diffuse supernovae flux search \cite{Olivares-DelCampo:2017feq} data as reanalyzed in Ref.~\cite{Arguelles:2019ouk} are shown for comparison.}
\label{fig:lmu_tau_limits_thermal}
\end{figure*}


In this model, after oscillations in transit from the Galactic Center, the flavor ratio of neutrinos on Earth is $\nu_{e}$:$\nu_{\mu}$:$\nu_{\tau}=1:2:2$. Hence for dark matter annihilation into $\mu$ and $\tau$ neutrinos, the electron (anti-)neutrino flux on Earth is given by
\begin{equation}
\frac{d\Phi}{dE_{\nu}} =\frac{1}{5} \frac{\langle\sigma v\rangle}{8\pi m_\chi^2} \frac{dJ}{d\Omega}  \delta(E_{\nu} - m_{\chi}).
\end{equation}
Here $\langle\sigma v\rangle$ is the thermally averaged annihilation cross-section in the present day:
\begin{equation}
\langle \sigma v\rangle = \frac{1}{N_{v}} \int_{0}^{\infty} \sigma v f(v) dv,
\end{equation}
where $N_{v} = \int_{0}^{\infty} f(v) dv$ and $f(v)$ is the Galactic velocity distribution, which we assume to be the Maxwell-Boltzmann distribution given by $f(v) = \sqrt{2/\pi} ~v^{2}/v_{0}^{3} ~e^{-v^{2}/2v_{0}^{2}}$, with $v_{0} = 220$ km/s. See Appendix~\ref{app:relic_dens} for the definition of $\langle \sigma v\rangle$ in the different dark matter models defined in Eq.~\eqref{eq:dm_models}. 

In Figure~\ref{fig:lmu_tau_limits_thermal} (left panel), we recast the projected model-independent limits derived in Section~\ref{sec:limits} to set model-specific constraints on dark matter annihilating into neutrinos from the Galactic Center. 
Here we assume benchmark parameters $m_{Z^{\prime}} = 3 m_{\chi}$, $g_{\chi} = 1$ and $g_{\mu-\tau} = 10^{-5}$. The orange shaded region is excluded by Borexino \cite{2011PhLB..696..191B}, the green shaded region is excluded by Kamland \cite{KamLAND:2011bnd,KamLAND:2021gvi}, magenta is bound by the Super-Kamiokande anti-neutrino analysis \cite{Super-Kamiokande:2015qek}, and the purple region is bound by a re-analysis of Super-Kamiokande atmospheric neutrino data \cite{Olivares-DelCampo:2017feq}. 
The grey horizontal dashed line represents the thermal annihilation cross-section corresponding to dark matter being in thermal equilibrium with the Standard Model bath at the time of decoupling.
The LArTPC detector projections are shown in the red and blue lines, assuming 40~kton$\times$year exposure. We show the projected reach both without directional information (full-sky, solid lines) and with directional information (dashed lines). The blue lines are the full-sky projections using both CC and ES events, while 
the red lines are the analyses using the ES-selected sample. 
We can see, as in Figure~\ref{fig:model_indep_sigmav}, that at lower masses directional information provides the strongest projections, however even without directional information, our assumed benchmark LArTPC can have the strongest projections at higher masses and reaches the thermal relic line, and would be competitive with existing limits after 40~kton$\times$year exposure assuming a directional analysis of ES-enriched events.\\


The black solid line in Figure~\ref{fig:lmu_tau_limits_thermal} (left) is the thermally averaged annihilation cross-section in the Dirac fermion dark matter case. The dashed and dot-dashed lines at the bottom of the figure are the Majorana and complex scalar cases respectively. The Dirac dark matter annihilation is $s$-wave dominated while the two latter cases are $p$-wave and hence velocity suppressed. As a result, their late-time annihilation cross-sections (for the benchmark values chosen) are small.
The vertical grey shaded region is the area in which dark matter annihilates during Big Bang Nucleosynthesis (BBN), increasing $N_{\rm eff}$ and disrupting the elemental abundances. This bound is obtained from both BBN and CMB data, but is somewhat dark matter model dependent (i.e., the bound can strengthen or weaken depending on the dark matter spin), here we have chosen the Dirac dark matter case, which is the strongest constraint \cite{Boehm:2012gr,Wilkinson:2016gsy,Escudero:2019gzq,Giovanetti:2021izc}.
In Figure~\ref{fig:lmu_tau_limits_thermal} (right panel), we convert the limits and projections discussed above into the $g_{\mu-\tau}$ vs.~$m_{Z^{\prime}}$ parameter space, with the benchmark parameters $m_{Z^{\prime}} = 3 m_{\chi}$ and $g_{\chi} = 1$. 
The shaded regions are as described previously. 
The black solid, dashed, and dot-dashed diagonal lines represents the points in parameter space where the relic abundance (described in Appendix~\ref{app:relic_dens}) matches the observed value of $\Omega h^{2} \sim 0.12$ \cite{Planck:2018vyg}, for the Dirac fermion, Majorana fermion and Complex scalar respectively.
The grey shaded parabolic region is excluded at $5 \sigma$ CL by measurement of the anomalous magnetic moment of the muon ($g$-2) \cite{Pospelov:2008zw}. The green shaded region bounded by a dashed band is the $2\sigma$ muon $g-2$ favored region given by the latest allowed range of $ \Delta a_{\mu}$, as reported by the Fermilab E989 experiment \cite{Muong-2:2021ojo}. The red line represents the current central value.
To focus on the large-volume neutrino detector projections and make the plot easier to read, we do not place the accelerator and astrophysical bounds (some of which overlap) here. We refer the reader to Refs.~\cite{Foldenauer:2018zrz,Bauer:2018onh}.

\section{Conclusions \label{sec:conclusions}}

Dark matter couplings to neutrinos are one of the least constrained possible interactions between the dark sector and the Standard Model. The low cross section of neutrinos serve to seclude such dark matter from experimental measurement; this is compounded at low energies, making limits here especially weak. 
At these low energies, a dominant contribution to the backgrounds for indirect detection of dark matter annihilations into neutrinos come from Solar neutrinos. The high level of directionality of these backgrounds, compared to the localized signal from the Galactic Center, raise the possibility that neutrino detectors that can resolve the path of scattered neutrinos may place stronger bounds on dark matter than otherwise would be possible.

In this paper, we have considered the prospects for future large-scale, deep underground liquid argon time projection chambers with directional reconstruction capabilities to search for this signal. We have shown that LArTPCs can be a powerful detector in this area, and will be most sensitive to dark matter with masses above $\sim$ 30 MeV, even without directional information. 
When including directionality, we find that a LArTPC with 40~kton$\times$year of exposure can set competitive limits on dark matter masses lower than $\sim$ 10 MeV, using both the directionality of the electrons induced from neutrino scattering and the ability to discriminate between charged current and elastic electron scattering events.  We have also illustrated the applicability of these results by considering an example model, showing that a DUNE-like LArTPC detector would be sensitive to thermally-produced Dirac fermion dark matter.

\section*{Acknowledgements}

We are grateful to Steven Gardiner, KC Kong, Gordan Krnjaic, Aaron Vincent, and Joseph Zennamo for valuable discussions. We also thank Ivan Lepetic for input on modeling the CC/ES discrimination. This work was initiated at the Aspen Center for Physics, which is supported by National Science Foundation grant PHY-1607611. This work was also partially supported by a grant from the Sloan Foundation.
MRB is supported by DOE grant DE-SC0017811. 
AM is supported by NSF grant PHY-2047665.
GM is supported by DOE grant DC-SC0012704 and by the Arthur B. McDonald Canadian Astroparticle Physics Research Institute. Research at Perimeter Institute is supported by the government of Canada through the Department of Innovation, Science and Economic Development and by the Province of Ontario through MEDJCT. GM also acknowledges support from the UC office of the President via the UCI Chancellor?s Advanced Postdoctoral Fellowship.\\

This work is based on the ideas and calculations of the authors, and publicly-available information pertaining the experimental capabilities. We speak on behalf of ourselves, and not on behalf of the DUNE Collaboration.

\appendix


\section{Differential $J$ Factor}
\label{app:jfactor}
The differential $J$ factor, appearing in the last term in Eq.~\eqref{eq:DMnuflux}, corresponds to a line-of-sight (l.o.s.) integral of the dark matter density squared (weighted by $1/\ell^2$), encoding the angular dependence of the signal. This term is given by
\begin{equation}
\frac{dJ}{d\Omega} = \int_{\rm l.o.s.} d\ell \rho_\chi^2(\ell,\Omega).
\end{equation}
As we are interested in the dark matter signal from annihilation in the Galactic Center (GC), we will consider a dark matter profile which is spherically symmetric, with 
\begin{eqnarray}
\rho_\chi(\ell,\Omega) & = & \rho_\chi(r) \\
r^2 & = & \ell^2+R_0^2 -2R_0 \ell \cos\theta_{\rm GC}. \nonumber
\end{eqnarray}
Here $R_0 = 8.112$~kpc is the distance between the Sun and the GC \cite{Abuter:2019aaa}, and  $\theta_{\rm GC}$ is the line-of-sight opening angle away from the Galactic Center. As a result of these assumptions, $dJ/d\Omega$ is independent of the azimuthal angle around the GC.

\begin{figure}[th]
\includegraphics[width=0.9\columnwidth]{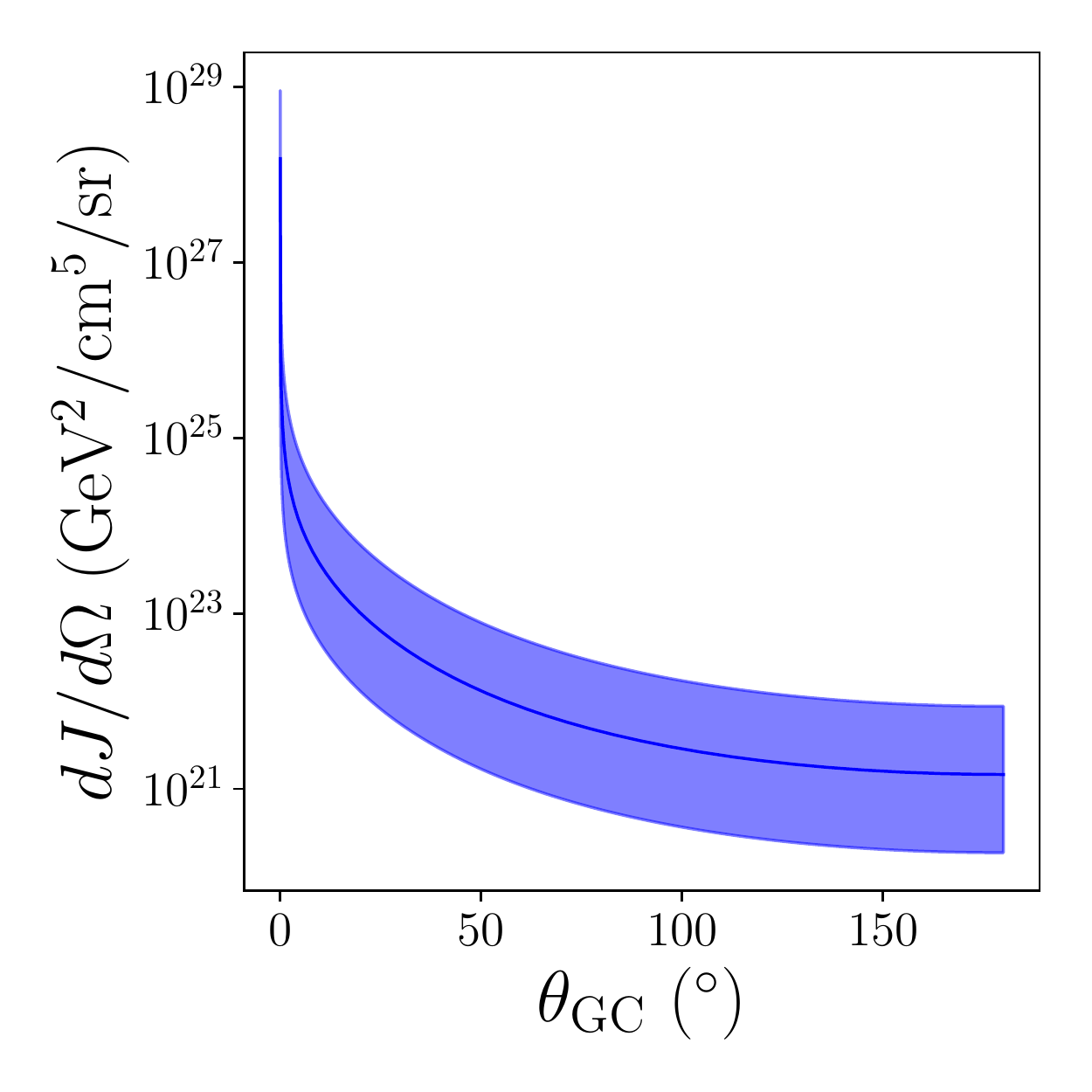}
\caption{Differential $J$-factor as a function of opening angle from the Galactic Center $dJ/d\theta$ assuming the best-fit NFW profile of Ref.~\cite{Fornasa:2013iaa}. Shaded region indicates $1\sigma$ error from the NFW profile fit. 
\label{fig:dJdOmega}}
\end{figure}

For the Galactic dark matter potential, we will use a Navarro-Frenk-White (NFW) \cite{Navarro:1995iw} profile
\begin{equation}
\rho_\chi(r) = \frac{\rho_0}{\left(\frac{r}{r_s} \right)\left(1+\frac{r}{r_s} \right)^2},
\end{equation}
with best-fit parameters \cite{Fornasa:2013iaa}
\begin{eqnarray}
\log_{10} \left(\rho_0/[M_\odot/{\rm pc}^3]\right) & = & -1.53^{+0.02}_{-0.68}  \\
r_s & = & \left(10.01^{+14.42}_{-0.12} \right)~{\rm kpc}. \nonumber
\end{eqnarray}
The differential $J$-factor as a function of $\theta_{\rm GC}$ is shown in Figure~\ref{fig:dJdOmega}. Integrated over the entire sky, this results in a $J$-factor of $\left(2.53^{+0.15}_{-1.19}\right)\times 10^{23}~{\rm GeV}^2/{\rm cm}^5$. We adopt the central value for our analysis, as variations in the differential $J$-factor within the $1\sigma$ errors of the fit parameters have a straightforward strengthening or weakening of projected limits on the annihilation cross section with very little change in angular dependence.

\section{Gauged Vector Portal Model}
\label{app:dmmodel}
In the gauged vector portal model we consider, the Lagrangian of interactions between Standard Model fermions $f$ and the dark matter $\chi$ is given by
\begin{equation}
\mathcal{L} \supset \frac{m_{Z^{\prime}}^{2}}{2} Z_{\mu}^{\prime} Z^{\prime \mu} + Z_{\mu}^{\prime} (g_{f} \mathcal{J}_{f}^{\mu} + \epsilon e \mathcal{J}_{EM}^{\mu} + g_{\chi} \mathcal{J}_{\chi}^{\mu}),
\end{equation}
where $m_{Z^{\prime}}$ is the gauge boson mass, $g_{f} \equiv Q_{\mu - \tau}^{f} g_{\mu - \tau}$ is the gauge coupling of the mediator to Standard Model fermions, and the currents ${\cal J}$ are defined below along with the kinetic mixing parameter $\epsilon$. For the Standard Model fermions, we assume unit charge: $Q_{\mu - \tau}^{f} \equiv 1$.

The $L_{\mu} - L_{\tau}$ current is given by 
\begin{equation}
\mathcal{J}_{f}^{\mu} = \bar{\mu} \gamma^{\mu} \mu + \bar{\nu}_{\mu} \gamma^{\mu} P_{L} \nu_{\mu} - \bar{\tau} \gamma^{\mu} \tau - \bar{\nu}_{\tau} \gamma^{\mu} P_{L} \nu_{\tau},
\end{equation} 
where $P_{L} = \frac{1}{2}(1 - \gamma^{5})$ is the left handed chirality operator.

In addition to dominant interactions of the $Z^\prime$ with $2^{\rm nd}$ and $3^{\rm rd}$ generation Standard Model leptons through a gauge coupling $g_{f}$, a one-loop level kinetic mixing between $Z^{\prime}$ and the neutral Standard Model gauge bosons is given by \cite{Escudero:2019gzq}:
\begin{equation}
\epsilon = - \frac{e g_{\mu - \tau}}{12 \pi^{2}} {\rm log} \left(\frac{m_{\tau}^{2}}{m_{\mu}^{2}} \right).
\end{equation}
All the other Standard Model fermions, represented by $\mathcal{J}_{EM}^{\mu} = \bar{f} \gamma^{\mu} f$, couple to $Z^{\prime}$ through kinetic mixing. We do not consider this subdominant interaction in our study.

Assuming that the dark matter $\chi$ is charged under this extra group, then the vector boson can mediate interactions between $\chi$ and the Standard Model, with the coupling of the mediator to $\chi \equiv Q_{\mu - \tau}^{\chi} g_{\mu - \tau}$. We assume that the dark matter is vector-like under this new gauge group, with a charge that can vary away from unity. Following the convention in Ref.~\cite{Kahn:2018cqs}, we choose three benchmark dark matter models where:
\begin{equation}
\mathcal{J}_{\chi}^{\mu} = \left\{\begin{array}{lc} \bar{\chi} \gamma^{\mu} \chi & {\rm Dirac ~Fermion} \\
\frac{1}{2} \bar{\chi} \gamma^{5} \gamma^{\mu} \chi & {\rm Majorana ~Fermion}\\
i \chi^{*} \partial^{\mu} \chi  &  {\rm Complex ~Scalar}
\end{array} \right.
\label{eq:dm_models}
\end{equation}

\section{Relic Density Calculations \label{app:relic_dens}}
To compute the relic density we begin with the thermally averaged annihilation cross-section, which shows up at both late and early times (in the relic density), as illustrated in fig.~\ref{fig:lmu_tau_limits_thermal}.
For the Dirac fermion dark matter case, the interaction Lagrangian is given by 
\begin{equation}
\mathcal{L} \supset g_{\chi} \overline{\chi} \gamma^{\mu} \chi Z^{\prime}_{\mu} + g_{\mu-\tau} \overline{f} \gamma^{\mu} f Z^{\prime}_{\mu} + g_{\mu-\tau} \overline{\nu}_{f} \gamma^{\mu} P_{L} \nu_{f} Z^{\prime}_{\mu},
\end{equation}
where $f = \{ \mu,\tau \}$. 
As an example, we focus on the $m_{Z^{\prime}} ~\textgreater ~m_{\chi} \textgreater ~m_{f}$ limit. 
Then for the annihilation process, $\bar{\chi} \chi \rightarrow \bar{f} f$, the cross-section in terms of Mandelstam s is given by
\begin{eqnarray}
\sigma_{\rm ann} (s) =  \frac{g_{\chi}^{2} g_{\mu-\tau}^{2}}{12 \pi s} \frac{\sqrt{s-4m_{f}^{s}} (s+2m_{f}^{2})(s+2m_{\chi}^{2})}{(s-4m_{\chi}^{2})[(s-m_{Z^{\prime}}^{2})^{2} + m_{Z^{\prime}}^{2} \Gamma_{Z^{\prime}}^{2}]}, 
\end{eqnarray}
where $\Gamma_{Z^{\prime}}$ is the decay width of the new $L_{\mu}-L_{\tau}$ gauge boson into $\mu$ and $\tau$ is given by
\begin{equation}
\Gamma_{Z^{\prime}} (Z^{\prime} \rightarrow \bar{f} f) = \frac{g_{\mu-\tau}^{2} m_{Z^{\prime}}}{12 \pi} \left(1 + \frac{2 m_{f}^{2}}{m_{Z^{\prime}}^{2}} \right) \sqrt{1 - \frac{4 m_{f}^{2}}{m_{Z^{\prime}}^{2}}}.
\end{equation}
The decay width into neutrinos is given by $\Gamma_{Z^{\prime}} = g_{\mu-\tau}^{2} m_{Z^{\prime}}/24 \pi$ \cite{Escudero:2019gzq, Kahn:2018cqs}.

To obtain the thermally averaged annihilation cross-section, we follow the formalism in Refs.~\cite{Wells:1994qy,Kahn:2018cqs}, and parametrize the annihilation cross-section as $\left< \sigma_{ann} v\right> \equiv \sigma_{0} x^{-n}$. The quantity $x \equiv m_{\chi}/T$, and n = 0 (for s-wave annihilation) and 1 (for p-wave annihilation).
Following the arguments above, we also obtain cross-sections for the Majorana and complex scalar dark matter cases. We generalize the cross-section for all three cases as
\begin{equation}
\sigma_{0} = \frac{g_{\mu-\tau}^{2} g_{\chi}^{2}}{k\pi m_{\chi}^{2}} \frac{(2 + m_{f}^{2}/m_{\chi}^{2})\sqrt{1-m_{f}^{2}/m_{\chi}^{2}}}{[(4-m_{Z^{\prime}}^{2}/m_{\chi}^{2})^{2} + m_{Z^{\prime}}^{2} \Gamma_{Z^{\prime}}^{2}/m_{\chi}^{4} ]},
\end{equation}
Following the thermal Freeze-out description in Ref.~\cite{Kahn:2018cqs}, we obtain the relic abundance of $\chi$ as
\begin{equation}
\Omega_{\chi} h^{2} = 8.77\times10^{-11} \frac{l(n+1)x_{f}^{n+1}{\rm GeV}^{-2}}{(g_{*,S}/\sqrt{g_{*}})\sigma_{0}},
\end{equation}
here $l$ accounts for dark matter degrees of freedom, $l=2$ if dark matter has an antiparticle and $l=1$ if it is its own antiparticle.
Here $g_{*,S}$ and $g_{*}$ are the entropic degrees of freedom and the relativistic degrees of freedom respectively. $x_{f}$ is the freeze-out temperature which can be solved recursively and is given by
\begin{equation}
x_{f} \approx {\rm ln}\left[ \frac{c(c+2)}{4\pi^{3}} \sqrt{\frac{45}{2}} \frac{g ~m_{\chi}~ m_{pl}}{\sqrt{g_{*}(m_{\chi}/x_{f})}} \frac{\sigma_{0} x_{f}^{-n}}{\sqrt{x_{f}} (1 - 3/2x_{f})}\right].
\end{equation}

The quantity c is obtained parametrically and here we use $c = 1/2$, g is the number of degrees of freedom and $m_{pl} = 1.22\times10^{19}$ GeV is the Planck mass.
For the Dirac, Majorana and complex scalar models $(n,k,l) = (0,2,2)$, (1,6,1) and (1,12,2) respectively.

\bibliography{neutrino}

\end{document}